\documentclass[prd,nofootinbib,12pt]{revtex4}
\usepackage{amssymb,amsmath,latexsym,braket,fancyref,hyphenat}
\usepackage[caption=false]{subfig}
\usepackage{graphicx}

\numberwithin{equation}{section}

\begin{document}

\begin{flushright}
CERN-TH-2021-162\\
October 2021
\end{flushright}

\title{{\Large Charged and CP-violating Kink Solutions\\[2mm]
    in  the Two Higgs Doublet Model}\\[3mm] }

\author {Kai Hong Law$^{\,a}\,$\footnote{E-mail address: {\tt kaihong.law@postgrad.manchester.ac.uk}} and Apostolos Pilaftsis$^{\,a,b}\,$\footnote{E-mail address: {\tt apostolos.pilaftsis@manchester.ac.uk}}}

\affiliation{\vspace{2mm}${}^a$Department of Physics and Astronomy, University of Manchester,\\ \vspace{2mm}Manchester,  M13 9PL, United Kingdom\\
${}$\vspace*{2mm} 
${}^b$Theoretical Physics Department, CERN, CH-1211 Geneva 23, Switzerland }

\begin{abstract}
\vspace{2mm}
\centerline{\small {\bf ABSTRACT} }
\vspace{2mm}

\noindent
Minimal extensions of the Standard Model (SM), such as the so-called Two Higgs Doublet Model (2HDM), can possess several accidental discrete symmetries whose spontaneous breakdown in the early universe usually lead to the formation of domain walls. We extend an earlier work~\cite{SDW} on this topic by studying in more detail the analytic properties of electrically charged and CP-violating kink solutions in the $Z_{2}$-symmetric 2HDM.  We derive the complete set of equations of motion that describe the 1D spatial profile of both the 2HDM vacuum parameters and the would-be Goldstone bosons $G^{1,2,3}$ of the SM. These equations are then solved numerically using the gradient flow technique, and the results of our analysis are presented in different parametrizations of the Higgs doublets${}$. In particular, we show analytically how an electrically charged profile should arise in 1D kink solutions\- when asymmetric boundary conditions are imposed on the Goldstone mode $G^{2}$ at spatial infinities, i.e.~as~${x\to \pm \infty}$. If~asymmetric boundary conditions are selected at~${x\to \pm \infty}$ for the Goldstone mode $G^{3}$ or the longitudinal mode~$\theta$ corresponding to a would-be massive photon, the derived kink solutions are then shown to exhibit CP violation. Possible cosmological implications of the electrically charged and CP-violating domain walls in the 2HDM are discussed.

\end{abstract}

\maketitle

\vfill\eject
\def\D{d}
\section{Introduction}

The Standard Model (SM) of particle physics has been tested in many different low-energy and collider experiments~\cite{UA1:1983crd,CDF:1995wbb,D0:1994wmk,Aad, Djouadi:2005gi}, and some of its predictions have been verified to a high degree of accuracy. Nevertheless, we still believe that the SM is not a complete theory, since it is unable to explain certain cosmological phenomena, such as the origin of dark matter and the matter-antimatter asymmetry in our universe. For the latter, CP violation is necessary to explain why there is more matter than antimatter in our universe~\cite{Sakharov:1967dj}. Although signatures of CP violation have been observed in particle physics experiments~\cite{Sachs:1964zz, BaBar:2004gyj, Belle:2004nch, LHCb:2019hro, LHCb:2020byh} in fairly good agreement with SM predictions, this CP violation in the electroweak sector is deemed to be insufficient to explain the observed Baryon Asymmetry in the Universe (BAU)~\cite{Gavela:1994dt, Farakos:1994xh}.

Many models that extend the particle content of the SM have been proposed in the literature, with the aim to address the cosmological problems mentioned above.  One such minimal and well studied extension of the SM is the so-called Two Higgs Doublet Model~(2HDM)~\cite{TDLee}. The 2HDM adds one more complex scalar doublet to the~SM, and so predicts the existence of five physical scalar particles, one of which can be identified with the SM Higgs boson which was observed at the LHC~\cite{Aad}. The 2HDM potential could provide new sources of CP violation~\cite{Pilaftsis:1999qt,Branco:2011iw,Carena:2015uoe,Bian:2016awe, Basler:2019iuu} that would be needed to account for the BAU~\cite{Cohen:1993nk}.  

There are several accidental symmetries that the 2HDM can acquire if certain parameter choices are met~\cite{Ivanov:2007de,Ferreira:2009wh,Battye:2011jj,Pilaftsis:2011ed}. The breaking of these symmetries can lead to topological defects in the model, such as domain walls, vortices and global monopoles~\cite{Battye:2011jj, Eto:2020opf,Eto:2020hjb}. The nature of the defect can be determined by the topology of the vacuum manifold~\cite{Kibble:1976sj}. Here we will focus on the discrete symmetry  $Z_{2}$, even though our approach can apply equally well to the other two discrete symmetries, such as the standard CP symmetry and its descendent symmetry CP2. Domain walls are formed when the $Z_{2}$ symmetry is broken during a phase transition of our universe. In this case, the vacuum manifold consists of disconnected regions of minima. During symmetry breaking, regions in space that are causally disconnected can fall into different minima of the potential. As a consequence, domains are formed and the boundary surfaces separating them are called domain walls.

Domain walls are of some concern because they can have detrimental cosmological implications\-.  In the early universe, discrete symmetries of a scalar potential are generically restored for sufficiently high temperatures. However, as the universe cools while possibly undergoing a series of symmetry breaking phase transitions, domain walls can form. Sub\-sequently, the energy density of the domain walls decreases as the universe expands. We may naively estimate the rate of decrease using a self-scaling argument. Within a Hubble radius $r$, the total energy of the domain walls is proportional to $Er^{2}$, where $E$ is the energy per unit area of a domain wall.  Then, the energy density of the domain wall $\rho_{\textrm{dw}}$, which is the energy per unit volume, follows the relation $\rho_{\textrm{dw}}\propto Er^{-1}$. Since the horizon expands at the speed of light, we have $\rho_{\textrm{dw}}\propto Et^{-1}$. Therefore, the energy density of domain walls scales as ${\textrm{(time)}}^{-1}$. However, the energy densities of matter and radiation scale down much faster as ${(\textrm{time})}^{-2}$ in their respective epochs \cite{Lazanu:2015fua}. Consequently, domain walls would grow relative to matter and radiation, and eventually dominate the energy density of the universe \cite{Press:1989yh, Larsson:1996sp}. Today we do not observe domain walls and as such, their absence indicates the possible existence of a mechanism, like inflation, or entails a specific choice of  model parameters that renders them harmless~\cite{Eto:2018hhg,Battye:2020jeu}.

Recently, it was found in numerical simulations~\cite{SDW} that 1D kink solutions obtained in 2HDMs may violate both the electric charge and CP, when asymmetric boundary conditions at spatial infinities, $x\to \pm \infty$, are imposed on the would-be Goldstone (longitudinal) modes, $G^{1,2,3}$ and $\theta$, related the SM gauge bosons, $W^\pm$ and $Z$, and to a would-be massive photon.  The asymmetric boundary conditions on the Goldstone modes represent a general pragmatic choice that one has to make in order to realistically describe the formation of domain walls and their evolution starting from initial random field configurations.

In this paper, we complement the earlier work of~\cite{SDW} by studying in more detail the analytic properties of charged and CP-violating kink solutions in the $Z_2$-symmetric 2HDM.  We first derive the equations of motion for all the parameters defining the vacuum manifold of the 2HDM, including the Goldstone modes. We use a non-linear representation of the two Higgs doublets, where the rotation angles are the would-be Goldstone bosons after electro\-weak symmetry breaking~\cite{Goldstone:1961eq}.  In particular, we show analytically how self-consistency of the kink solutions with asymmetric boundary conditions for the Goldstone modes necessarily implies the occurence of charged and CP-violating domain walls. These findings are confirmed by solving numerically the pertinent equations of motion using the gradient flow technique, and they are in good agreement with the earlier study in~\cite{SDW}.  

The present article is organised as follows. After this introductory section, in Section~\ref{Chapter4} we present a brief discussion of the 2HDM, including different parameterisations of the two Higgs doublets, along with the mass matrices of the physical scalars. In Section~\ref{Chapter6}, we consider the most general parametrization of the two Higgs doublets by means of an electroweak gauge transformation and thus allow for the possible presence of charge-breaking and CP-violating vacua. In addition, we study analytically all the kink solutions for each Goldstone mode individually, and verify our findings by solving numerically the pertinent equations of motion using the gradient flow method. Basic aspects of the gradient flow method are reviewed in Appendix~\ref{AppendixA}.  Finally, Section \ref{Chapter7} summarises our results and discusses possible cosmological phenomena due to the electrically charged and CP-violating domain walls that may take place in the 2HDM and beyond.

\section{The $Z_{2}$-symmetric Two Higgs Doublet Model}
\label{Chapter4}

The scalar potential of the  $Z_{2}$-symmetric 2HDM reads
\begin{equation}\label{eq16}
\begin{split}
V(\Phi_{1}, \Phi_{2})=&-\mu_{1}^{2}(\Phi_{1}^{\dagger}\Phi_{1})-\mu_{2}^{2}(\Phi_{2}^{\dagger}\Phi_{2})+\lambda_{1}{(\Phi_{1}^{\dagger}\Phi_{1})}^{2}+\lambda_{2}{(\Phi_{2}^{\dagger}\Phi_{2})}^{2}+\lambda_{3}(\Phi_{1}^{\dagger}\Phi_{1})(\Phi_{2}^{\dagger}\Phi_{2}) \\
& +(\lambda_{4}-|\lambda_{5}|){[\textrm{Re}(\Phi_{1}^{\dagger}\Phi_{2})]}^{2}+(\lambda_{4}+|\lambda_{5}|){[\textrm{Im}(\Phi_{1}^{\dagger}\Phi_{2})]}^{2}\, .
\end{split}
\end{equation}
Obviously, the potential as defined in~\eqref{eq16} is invariant under the $Z_{2}$ symmetry~\cite{Glashow:1976nt}, 
\begin{equation}
\Phi_{1}\rightarrow \Phi_{1}\:,\qquad \Phi_{2}\rightarrow -\,\Phi_{2}\; .
\end{equation}

\subsection{Parametrizations of the Higgs Doublets}

There are different ways to parametrize or represent the Higgs doublets in the 2HDM. The simplest representation is the linear one:
\begin{equation}
    \label{eq:linrep}
  \Phi_{1}=\begin{pmatrix}\phi_{1}+i\phi_{2} \\ \phi_{3}+i\phi_{4} \end{pmatrix},\qquad \Phi_{2}=\begin{pmatrix}\phi_{5}+i\phi_{6} \\ \phi_{7}+i\phi_{8}
  \end{pmatrix}\, .
\end{equation}
Another equivalent and perhaps more intuitive parametrization to interprete our results in Section~\ref{Chapter6} is the non-linear representation, which employs
an electroweak gauge trans\-formation and so renders the potential occurence of
charge- and CP-breaking vacua manifest${}$~\cite{SDW}. Such vacua are
in general admissible in the 2HDM~\cite{Branco:2011iw} and may be expressed in terms of the four vacuum parameters $v_{1}, v_{2}, v_{+}$ and $\xi$ as follows:
\begin{equation}\label{eq4.9}
\Phi_{1}^{0}=\frac{1}{\sqrt{2}}\begin{pmatrix}0 \\ v_{1}\end{pmatrix},\qquad \Phi_{2}^{0}=\frac{1}{\sqrt{2}}\begin{pmatrix}v_{+} \\ v_{2}e^{i\xi}\end{pmatrix}\, .
\end{equation}
If $v_1$ is non-zero, then a non-vanishing value for $v_{+}$ implies that the ground state $(\Phi^0_{1}\,,\,\Phi^0_2)$ is  charge-violating or electrically charged, while a relative non-zero phase $\xi$ (possibly not a multiple of $\pi/2$~\cite{Branco:2011iw}) between the two neutral vacuum parameters $v_{1,2}$ implies that the ground state violates CP. To allow for the most general vacuum field configurations beyond the unitary gauge, we parametrize the Higgs doublets $\Phi_{1,2}$ non-linearly by means of an $\text{SU(2)}_{L}\times\textrm{U(1)}_{Y}$ gauge transformation,
\begin{equation}
\label{eq8}
\Phi_1\, =\, U\,\Phi_{1}^{0}\,,\qquad  \Phi_2\, =\, U\,\Phi_{2}^{0}\,,
\end{equation}
where 
\begin{equation}
\label{eq9}
U\, =\, e^{i\theta}\exp\left(i\frac{G^{a}}{v_{\textrm{SM}}}\frac{\sigma^{a}}{2}\right)\, =\, e^{i\theta}\exp\left(\frac{i\widehat{G}^{a}\sigma^{a}}{2}\right)
\end{equation}
is an element of the $\text{SU(2)}_{L}\times\textrm{U(1)}_{Y}$ gauge group. In~\eqref{eq9}, $\theta$ and $G^{a}=(G^{1}, G^{2}, G^{3})$ (with $\widehat{G}^a \equiv G^a/v_{\rm SM}$) are the would-be Goldstone bosons after electroweak symmetry breaking, and ${v_{\textrm{SM}} \simeq 246\:\textrm{GeV}}$ is the vacuum expectation value (VEV) of the SM Higgs doublet. In this parametrization, we may easily count that we have eight parameters in total to represent the vacua, i.e.~$v_{1,2},\, \xi,\, v_+,\, G^a$ and $\theta$,  which is the same number of parameters as in the linear representation~\eqref{eq:linrep}.

On the other hand, using the bilinear scalar-field formalism \cite{Nishi:2006tg,Maniatis:2006fs,Ivanov:2007de}, we may introduce a four-vector $R^{\mu}$
which is invariant under electroweak gauge transformations:
\begin{equation}
R^{\mu}\equiv\Phi^{\dagger}\sigma^{\mu}\Phi=\begin{pmatrix}\Phi_{1}^{\dagger}\Phi_{1}+\Phi_{2}^{\dagger}\Phi_{2} \\ \Phi_{1}^{\dagger}\Phi_{2}+\Phi_{2}^{\dagger}\Phi_{1} \\ -i[\Phi_{1}^{\dagger}\Phi_{2}-\Phi_{2}^{\dagger}\Phi_{1}] \\ \Phi_{1}^{\dagger}\Phi_{1}-\Phi_{2}^{\dagger}\Phi_{2} \end{pmatrix}\:,
\end{equation}
where $\Phi={(\Phi_{1}, \Phi_{2})}^{\sf T}$. The index $\mu$ in $\sigma^{\mu}$ runs from 0 to 3, with $\sigma^{0}=I_{2}$ and $\sigma^{1,2,3}$ being the Pauli matrices.  In terms of $R^{\mu}$, the general 2HDM potential may be written as
\begin{equation}\label{eq53}
V=-\frac{1}{2}M_{\mu}R^{\mu}+\frac{1}{4}L_{\mu\nu}R^{\mu}R^{\nu}\:,
\end{equation}
where
\begin{equation}
M_{\mu}=\begin{pmatrix}\mu_{1}^{2}+\mu_{2}^{2}\:, & 2\textrm{Re}(m_{12}^{2})\:, & -2\textrm{Im}(m_{12}^{2})\:, & \mu_{1}^{2}-\mu_{2}^{2} \end{pmatrix}\:,
\end{equation}
\begin{equation}\label{eq4.15}
L_{\mu\nu}=\begin{pmatrix}\lambda_{123} & \textrm{Re}(\lambda_{67}) & -\textrm{Im}(\lambda_{67}) & \bar{\lambda}_{12} \\ \textrm{Re}(\lambda_{67}) & \lambda_{4}+\textrm{Re}(\lambda_{5}) & -\textrm{Im}(\lambda_{5}) & \textrm{Re}(\bar{\lambda}_{67})\\-\textrm{Im}(\lambda_{67}) & -\textrm{Im}(\lambda_{5}) & \lambda_{4}-\textrm{Re}(\lambda_{5}) & -\textrm{Im}(\bar{\lambda}_{67})\\\bar{\lambda}_{12} & \textrm{Re}(\bar{\lambda}_{67}) & -\textrm{Im}(\bar{\lambda}_{67}) & \bar{\lambda}_{123}\end{pmatrix}\,.
\end{equation}
The first term in \eqref{eq53} contains the mass terms, while the second term describes the quartic couplings. In \eqref{eq4.15}, we have used the notations: $\lambda_{ab}=\lambda_{a}+\lambda_{b}$, $\bar{\lambda}_{ab}=\lambda_{a}-\lambda_{b}$, $\lambda_{abc}=\lambda_{a}+\lambda_{b}+\lambda_{c}$ and $\bar{\lambda}_{abc}=\lambda_{a}+\lambda_{b}-\lambda_{c}$. Since $R^{\mu}$ is invariant under the unitary transformations~$U$ of the SM gauge group, we can use the charge-breaking
ground state $(\Phi^0_1\,,\,\Phi^0_2)$ (in the unitary gauge) to express this four-vector
as follows:
\begin{equation}\label{eq4.1.9}
R^{\mu}=\frac{1}{2}\begin{pmatrix}v_{1}^{2}+v_{2}^{2}+v_{+}^{2} \\ 2v_{1}v_{2}\cos\xi \\ 2v_{1}v_{2}\sin\xi \\ v_{1}^{2}-v_{2}^{2}-v_{+}^{2} \end{pmatrix}\,.
\end{equation}
Inverting the relations in \eqref{eq4.1.9}, we can express the vacuum parameters in terms of the components $R^{\mu}$. In particular, $v_{+}$ is related to the components $R^{\mu}$ by
\begin{equation}
v_{+}^{2}=\frac{R_{\mu}R^{\mu}}{R^{0}+R^{3}}\, .
\end{equation}
Therefore, we can determine whether a solution is charge-violating by looking at either the parameter $v_{+}$, when  for $v_1\neq 0$, or the norm of the 4-vector $R^{\mu}$. Imposing the condition $R_{\mu}R^{\mu}= v^2_1 v^2_+=0$, known as the vacuum neutrality condition~\cite{Ivanov:2007de}, would imply that $v_{+}=0$ or $v_1 = 0$. 

However, in order to properly describe the dynamics emerging from the Goldstone mode~$\theta$, we have to extend the above bilinear formalism and promote $R^{\mu}$ to an $\textrm{SU(2)}_{L}$-invariant six-vector $R^{A}$~\cite{Battye:2011jj, Pilaftsis:2011ed}, \begin{equation}\label{eq4.1.11}
  R^{A}=\begin{pmatrix}\Phi_{1}^{\dagger}\Phi_{1}+\Phi_{2}^{\dagger}\Phi_{2} \\ \Phi_{1}^{\dagger}\Phi_{2}+\Phi_{2}^{\dagger}\Phi_{1} \\ -i[\Phi_{1}^{\dagger}\Phi_{2}-\Phi_{2}^{\dagger}\Phi_{1}] \\ \Phi_{1}^{\dagger}\Phi_{1}-\Phi_{2}^{\dagger}\Phi_{2} \\ \Phi_{1}^{T}i\sigma^{2}\Phi_{2}-\Phi_{2}^{\dagger}i\sigma^{2}\Phi_{1}^* \\ -i[\Phi_{1}^{T}i\sigma^{2}\Phi_{2}+\Phi_{2}^{\dagger}i\sigma^{2}\Phi_{1}^*]\end{pmatrix}=\frac{1}{2}\begin{pmatrix}v_{1}^{2}+v_{2}^{2}+v_{+}^{2} \\ 2v_{1}v_{2}\cos\xi \\ 2v_{1}v_{2}\sin\xi \\ v_{1}^{2}-v_{2}^{2}-v_{+}^{2} \\ -2v_{1}v_{+}\cos 2\theta \\ -2v_{1}v_{+}\sin 2\theta
  \end{pmatrix}\,.
\end{equation}
Note that $R^{A}$ is a null vector, and its U(1)$_Y$-violating components $R^{4,5}$
do explicitly depend on $\theta$ which would correspond to a massive photon
for possible non-zero values of $v_+$.

To fully cover the parameter space of the vacuum manifold, we may choose $R^{0}, R^{1}, R^{2}, R^{3}$ and $R^{4}$, together with $G^{1}, G^{2}$ and $G^{3}$, as our free vacuum parameters. Evidently, these are eight independent quantities, in agreement with the total number of parameters needed to parametrise the vacuum manifold in the linear representation [cf.~\eqref{eq:linrep}].

\subsection{Mass Matrices}

In the 2HDM, we have five physical Higgs states, which in the absence of CP violation, the two scalars, $h$ and $H$, are CP-even, one is CP-odd,~$A$, and the remaining two scalars~$H^{\pm}$ are electrically charged. To find the mass matrices, it would be useful to use the following representation of the Higgs doublets:
\begin{equation}\label{eq4.2.1}
\Phi_{1}=\begin{pmatrix}\phi_{1}^{+} \\ \frac{1}{\sqrt{2}}\big(v_{1}+\phi_{1}+ia_{1}\big) \end{pmatrix}\:,\qquad \Phi_{2}=e^{i\xi}\begin{pmatrix}\phi_{2}^{+} \\ \frac{1}{\sqrt{2}}\big(v_{2}+\phi_{2}+ia_{2}\big) \end{pmatrix}\:,
\end{equation}
where $\phi_{1}^{+}$ and $\phi_{2}^{+}$ are complex scalar fields. $\phi_{1}$ and $\phi_{2}$ can be expressed as a rotation of the CP-even fields $h$ and $H$, while $a_{1}$ and $a_{2}$ can be expressed as a rotation of the CP-odd fields $G^{0}$ and $A$. More explicitly, we have
\begin{equation}
\begin{pmatrix}\phi_{1} \\ \phi_{2}\end{pmatrix}=\begin{pmatrix} c_{\alpha} & -s_{\alpha} \\ s_{\alpha} & c_{\alpha} \end{pmatrix}\begin{pmatrix} h \\ H\end{pmatrix}\:,\:\:\:\begin{pmatrix}a_{1} \\ a_{2}\end{pmatrix}=\begin{pmatrix} c_{\beta} & -s_{\beta} \\ s_{\beta} & c_{\beta} \end{pmatrix}\begin{pmatrix} G^{0} \\ A\end{pmatrix}\:,\end{equation}
where the short-hand notations $s_x\equiv\sin x$ and $c_x\equiv\cos x$ are employed for
the trigonometric functions. We note that $G^{0}$ is the would-be Goldstone boson
associated with the longitudinal polarization of the $Z$ boson. Since the $Z_{2}$-symmetric 2HDM potential is CP-preserving,  we may set $\xi=0$~\cite{SDW} to a good approximation, with possible exceptions arising  from instanton effects~\cite{Battye:2020jeu}. Substituting the representations of the Higgs doublets in~\eqref{eq4.2.1} back into the $Z_{2}$-symmetric potential in \eqref{eq16}, the CP-even, CP-odd and charged scalar mass matrices
are found respectively to be
\begin{eqnarray}
  \label{eq4.2.3}
  \mathcal{M}_{h,H}^{2} &=&
\left<\frac{\partial^{2}V}{\partial\phi_{i}\partial\phi_{j}}\right>\, =\, \begin{pmatrix}2\lambda_{1}v_{1}^{2} & \tilde{\lambda}_{345}v_{1}v_{2} \\ \tilde{\lambda}_{345}v_{1}v_{2} & 2\lambda_{2}v_{2}^{2} \end{pmatrix}\,,\\
  \mathcal{M}_{A}^{2} &=& \left<\frac{\partial^{2}V}{\partial a_{i}\partial a_{j}}\right>
                          \, =\,|\lambda_{5}|\begin{pmatrix}v_{2}^{2} & -v_{1}v_{2} \\ -v_{1}v_{2} & v_{1}^{2} \end{pmatrix}\,,\\
\mathcal{M}_{H^{\pm}}^{2} &=& \left<\frac{\partial^{2}V}{\partial \phi_{i}^{+}\partial \phi_{j}^{-}}\right>\, =\,-\frac{1}{2}(\lambda_{4}-|\lambda_{5}|)\begin{pmatrix}v_{2}^{2} & -v_{1}v_{2} \\ -v_{1}v_{2} & v_{1}^{2} \end{pmatrix}\,.
\end{eqnarray}
The squared masses of the five physical scalars, $h,\ H,\ A$ and $H^\pm$, are then given by
\begin{eqnarray}
  M_{h}^{2} &=&
\lambda_{1}v_{1}^{2}+\lambda_{2}v_{2}^{2}-\sqrt{{(\lambda_{1}v_{1}^{2}-\lambda_{2}v_{2}^{2})}^{2}+\tilde{\lambda}_{345}^{2}v_{1}^{2}v_{2}^{2}}\:,\\
M_{H}^{2} &=& \lambda_{1}v_{1}^{2}+\lambda_{2}v_{2}^{2}+\sqrt{{(\lambda_{1}v_{1}^{2}-\lambda_{2}v_{2}^{2})}^{2}+\tilde{\lambda}_{345}^{2}v_{1}^{2}v_{2}^{2}}\:,\\
    \label{eq4.2.8}
M_{A}^{2} &=&|\lambda_{5}|v_{\textrm{SM}}^{2}\:,\\
   \label{eq4.2.9}
M_{H^{\pm}}^{2} &=& -\frac{1}{2}(\lambda_{4}-|\lambda_{5}|)v_{\textrm{SM}}^{2}\:,
\end{eqnarray}
with $\tilde{\lambda}_{345}\equiv\lambda_{3}+\lambda_{4}-|\lambda_{5}|$. The VEVs of the Higgs doublets are related to the SM VEV $v_{\textrm{SM}}$ by the mixing angle $\beta$
that enters the diagonalisation of CP-odd scalar matrix, i.e.
\begin{equation}\label{eq4.2.10}
v_{1}^{0}\, =\, c_{\beta}\,v_{\textrm{SM}}\:,\qquad v_{2}^{0}\, =\, s_{\beta}\,v_{\textrm{SM}}\, .
\end{equation}
As done in~\cite{SDW}, in all our numerical simulations, we choose the masses of all the new heavy scalars to be equal: $M_{H}=M_{A}=M_{H^{\pm}}=200~\textrm{GeV}$, $\tan\beta=0.85$, and $\cos(\alpha-\beta)=1$ obeying the alignment limit, as this is dictated by a maximally symmetric Sp(4) realization of the 2HDM~\cite{BhupalDev:2014bir,Darvishi:2019ltl}.
Moreover,  we adopt the Type-I pattern of Yukawa interactions to avoid the majority of the phenomenological quark-flavour constraints.

\section{Electrically Charged and CP-violating Kink Solutions}
\label{Chapter6}

In a 1D spatial approximation, e.g.~along the $x$-direction, the total energy density of the electroweak gauged $Z_{2}$-symmetric 2HDM may be conveniently determined as~\cite{SDW} 
\begin{equation}\label{eq6.0.4}
\mathcal{E}\, =\, \frac{d\Phi_{1}^{\dagger}}{dx}\frac{d\Phi_{1}}{dx}+\frac{d\Phi_{2}^{\dagger}}{dx}\frac{d\Phi_{2}}{dx}+V(\Phi_{1}, \Phi_{2})\,, 
\end{equation}
where the scalar potential $V(\Phi_{1}, \Phi_{2})$ is defined in~\eqref{eq16}.  If we now use the non-linear representation as given in~\eqref{eq8} for the two Higgs doublets, i.e.~$\Phi_{1,2} = U \Phi^0_{1,2}$, we then observe that the 2HDM scalar potential simplifies as $V(\Phi_{1}, \Phi_{2}) = V(\Phi^0_{1}, \Phi^0_{2})$, and all Goldstone modes $G^{1,2,3}$ and $\theta$ contained in the unitary matrix $U$ [cf.~\eqref{eq9}] vanish identically.

To further simplify matters, we consider that only one normalized Goldstone mode~$\widehat{G}^{a} \equiv G^a/v_{\rm SM}$ is non-zero each time of our analytical investigation and may possess asymmetric boundary conditions at spatial infinities as $x \to \pm\infty$.  This corresponds to choosing a fixed given axis for performing an $\textrm{SU}(2)_L$ gauge rotation. With this simplification, we have
\begin{equation}
  \label{eq6.0.6} \frac{dU}{dx}\ =\ i\left(\frac{d\theta}{dx}+\frac{d\widehat{G}^{a}}{dx}\frac{\sigma^{a}}{2}\right)U\:, \end{equation}
where we reiterate that the index $a$ is {\em not} summed over. Substituting \eqref{eq8} into \eqref{eq6.0.4}, and using \eqref{eq6.0.6}, we find that the kinetic part of the energy density can be expressed as
\begin{equation}
   \label{eq11old}
  \begin{split}    \mathcal{E}_{\textrm{kin}}=&{\left\lvert\frac{d\Phi_{i}^{0}}{dx}\right\rvert}^{2}+{|\Phi_{i}^{0}|}^{2}\left[\left(\frac{d\theta}{dx}\right)^{2}+\frac{1}{4}\left(\frac{d\widehat{G}^{a}}{dx}\right)^{2}\right]+\Phi_{i}^{0\dagger}U^{\dagger}\left(\frac{d\theta}{dx}\right)\left(\frac{d\widehat{G}^{a}}{dx}\right)\sigma^{a}U\Phi_{i}^{0}\\    &+\left[i\frac{d\Phi_{i}^{0\dagger}}{dx}U^{\dagger}\left(\frac{d\theta}{dx}+\frac{d\widehat{G}^{a}}{dx}\frac{\sigma^{a}}{2}\right)U\Phi_{i}^{0}+\textrm{H.c.}\right]\,,
\end{split}
\end{equation}
where summation over the index $i =1,2$ is implied.
Since the unitary matrix $U$ in~\eqref{eq11old} involves the exponentiation of only one Pauli matrix $\sigma^{a}$, it commutes with the Pauli matrix $\sigma^{a}$ itself. Hence, $\mathcal{E}_{\textrm{kin}}$ will take on the simpler form 
\begin{equation}
  \label{eq11} \begin{split}    \mathcal{E}_{\textrm{kin}}=&{\left\lvert\frac{d\Phi_{i}^{0}}{dx}\right\rvert}^{2}+{|\Phi_{i}^{0}|}^{2}\left[\left(\frac{d\theta}{dx}\right)^{2}+\frac{1}{4}\left(\frac{d\widehat{G}^{a}}{dx}\right)^{2}\right]+\Phi_{i}^{0\dagger}\left(\frac{d\theta}{dx}\right)\left(\frac{d\widehat{G}^{a}}{dx}\right)\sigma^{a}\Phi_{i}^{0}\\    &+\left[i\frac{d\Phi_{i}^{0\dagger}}{dx}\left(\frac{d\theta}{dx}+\frac{d\widehat{G}^{a}}{dx}\frac{\sigma^{a}}{2}\right)\Phi_{i}^{0}+\textrm{H.c.}\right]\,.\end{split}
\end{equation}

\subsection{Goldstone Bosons}

Considering the simplified energy density $\mathcal{E}_{\textrm{kin}}$ in~\eqref{eq11} where each time only one of the would-be Goldstone bosons~$\widehat{G}^{a}$ is non-zero, we can derive a simpler set of equations for $\widehat{G}^{a}$,
\begin{equation}\label{eq18}
\frac{d}{dx}\biggl[\,\frac{1}{2}{|\Phi_{i}^{0}|}^{2}\frac{d\widehat{G}^{a}}{dx}+\Phi_{i}^{0\dagger}\sigma^{a}\Phi_{i}^{0}\frac{d\theta}{d x} +
\biggl(\frac{i}{2}\frac{d\Phi_{i}^{0\dagger}}{dx}\sigma^{a}\Phi_{i}^{0}+\textrm{H.c.}\biggr)\biggr]\ =\ 0
\end{equation}
Equation~\eqref{eq18} implies that the terms inside the derivative must add up to an $x$-independent constant. Since the $x$-derivatives of all the vacuum parameters 
should vanish as $x\to \pm\infty$ to ensure that the total energy of the kink solution 
is finite, this integration constant can only be zero.
In this way, we arrive at a first order differential equation describing the spatial profiles of $\widehat{G}^{a}$,
\begin{equation}
  \label{eq:NoetherGa}
\frac{1}{2}{|\Phi_{i}^{0}|}^{2}\frac{d\widehat{G}^{a}}{dx}+\Phi_{i}^{0\dagger}\sigma^{a}\Phi_{i}^{0}\frac{d\theta}{d x}+\biggl(\frac{i}{2}\frac{d\Phi_{i}^{0\dagger}}{dx}\sigma^{a}\Phi_{i}^{0}+\textrm{H.c.}\biggr)
\ =\ 0\,.
\end{equation}
By analogy, a similar first-order differential equation may be derived for $\theta(x)$, 
\begin{equation}
  \label{eq:NoetherTheta}
2{|\Phi_{i}^{0}|}^{2}\frac{d\theta}{dx}+\Phi_{i}^{0\dagger}\sigma^{a}\Phi_{i}^{0}\frac{d\widehat{G}^{a}}{dx}+\biggl(i\frac{d\Phi_{i}^{0\dagger}}{dx}\Phi_{i}^{0}+\textrm{H.c.}\biggr)=0\, .
\end{equation}
We note that the first-order differential equations in~\eqref{eq:NoetherGa} and~\eqref{eq:NoetherTheta} reflect the conservation of the
Noether currents associated with the local symmetries of the original theory under the  SU(2)$_L\times$U(1)$_Y$ group~\cite{SDW}.

In the following, we will derive the equations of motion for all the vacuum parameters by 
assuming that only one would-be Goldstone boson at the time, i.e.~$\theta (x)$ or $\widehat{G}^{1,2,3}$, is non-zero. Such a simplification enables us to better understand the analytic properties of the kink solutions when asymmetric boundary conditions are imposed on each of the would-be Goldstone bosons. In tandem, we use the gradient flow technique to obtain numerical solutions which will then be compared with our analytical findings.
A brief introduction to the gradient flow technique is given in Appendix~\ref{AppendixA}.

\subsubsection{The $\theta$-Scenario}
\label{Subsection6.1.1}

We start by considering the case where only $\theta (x)$ is non-zero, with all $G^{1,2,3}(x) = 0$. For brevity, we call this the $\theta$-scenario.  This means that the unitary matrix $U$ in~\eqref{eq9} describing an arbitrary gauge rotation is simply $U=e^{i\theta(x)}{\bf 1}_2$. In this case, the kinetic energy density of the system becomes \begin{equation} \begin{split}
    \mathcal{E}_\textrm{kin}\ =&\ \frac{1}{2}{\left(\frac{dv_{1}}{dx}\right)}^{2}+\frac{1}{2}{\left(\frac{dv_{2}}{dx}\right)}^{2}+\frac{1}{2}{\left(\frac{dv_{+}}{dx}\right)}^{2}+\frac{1}{2}v_{2}^{2}{\left(\frac{d\xi}{dx}\right)}^{2}\\
    &\ +\,\frac{1}{2}(v_{1}^{2}+v_{2}^{2}+v_{+}^{2}){\left(\frac{d\theta}{dx}\right)}^{2}+v_{2}^{2}\frac{d\xi}{dx}\frac{d\theta}{dx}\; .  \end{split} \end{equation} The gradient flow equations for $v_1, v_2, \xi, v_{+}$ and $\theta$ are then found to be \begin{equation}
    \label{eq:FlowTheta}
\begin{aligned}
\frac{\partial v_{1}}{\partial t}&=\frac{{\partial}^{2}v_{1}}{\partial x^{2}}-v_{1}\left(\frac{\partial\theta}{\partial x}\right)^{2}+\mu_{1}^{2}v_{1}-\lambda_{1}v_{1}^{3}-\frac{1}{2}\lambda_{3}v_{1}v_{+}^{2}-\frac{1}{2}(\lambda_{34}-|\lambda_{5}|c_{2\xi})v_{1}v_{2}^{2},\\
\frac{\partial v_{2}}{\partial t}&=\frac{{\partial}^{2}v_{2}}{\partial x^{2}}-v_{2}\left[{\left(\frac{\partial\xi}{\partial x}\right)}^{2}+\left(\frac{\partial\theta}{\partial x}\right)^{2}+2\frac{\partial\xi}{\partial x}\frac{\partial\theta}{\partial x}\right]
+\mu_{2}^{2}v_{2}-\lambda_{2}v_{2}(v_{2}^{2}+v_{+}^{2})-\frac{1}{2}(\lambda_{34}-|\lambda_{5}|c_{2\xi})v_{1}^{2}v_{2}\,,\\
\frac{\partial v_{+}}{\partial t}&=\frac{\partial^{2} v_{+}}{\partial x^{2}}-v_{+}{\left(\frac{\partial\theta}{\partial x}\right)}^{2}+\mu_{2}^{2}v_{+}-\lambda_{2}v_{+}(v_{2}^{2}+v_{+}^{2})-\frac{1}{2}\lambda_{3}v_{1}^{2}v_{+}\,,\\[2mm]
\frac{\partial\xi}{\partial t}&=v_{2}^{2}\left(\frac{{\partial}^{2}\xi}{\partial x^{2}}+\frac{\partial^{2}\theta}{\partial x^{2}}\right)+2v_{2}\frac{\partial v_{2}}{\partial x}\left(\frac{\partial\xi}{\partial x}+\frac{\partial\theta}{\partial x}\right)-\frac{1}{2}|\lambda_{5}|v_{1}^{2}v_{2}^{2}s_{2\xi}\,,\\[2mm]
\frac{\partial\theta}{\partial t}&=(v_{1}^{2}+v_{2}^{2}+v_{+}^{2})\left(\frac{\partial^{2}\theta}{\partial x^{2}}\right)+v_{2}^{2}\frac{\partial^{2}\xi}{\partial x^{2}}+2\frac{\partial\theta}{\partial x}\left(v_{1}\frac{\partial v_1}{\partial x}+v_{2}\frac{\partial v_2}{\partial x}+v_{+}\frac{\partial v_+}{\partial x}\right)+2v_{2}\frac{\partial v_2}{\partial x}\frac{\partial\xi}{\partial x}\,,
\end{aligned}
\end{equation}
where $t$ represents a fictitious time upon which each vacuum parameter is assumed to depend in this gradient flow method. A solution describing the ground state of the system is declared to be found, once the left-hand sides (LHSs) of the partial differential equations in~\eqref{eq:FlowTheta} will all vanish, up to a given degree of numerical accuracy. More details of the gradient flow method are given in Appendix~\ref{AppendixA}.

Assuming that the derivatives of all Goldstone bosons and vacuum parameters tend to zero at the boundaries as $x\to \pm \infty$, the gradient flow equation for $\theta(x)$
in the ground state will be
\begin{equation}
  (v_{1}^{2}+v_{2}^{2}+v_{+}^{2})\,\frac{d\theta}{dx}\: +\:
  v_{2}^{2}\,\frac{d\xi}{d x}\ =\ 0\, .
\end{equation}
Solving this last equation for $d\theta/dx$ yields
\begin{equation}
\label{eq3.10}
\frac{d\theta}{d x}\: =\: -\,\frac{v_{2}^{2}}{v_{1}^{2}+v_{2}^{2}+v_{+}^{2}}\,
\frac{d\xi}{d x}\, .
\end{equation}
Equation \eqref{eq3.10} is one of the central results of this paper. It tells us that if asymmetric boundary conditions at infinity are imposed on $\theta (x)$, i.e.~$\theta (-\infty) \neq \theta (+\infty)$, so that $d\theta/dx$ happens to be non-zero for a finite $x$-interval, then one must necessarily have $d\xi/dx \neq 0$ for a correlated $x$-interval of finite size, provided $v_2 \neq 0$ in the same interval. As expected from earlier considerations where the effect of Goldstone bosons was ignored~\cite{Battye:2011jj}, this is indeed the case, so the 1D kink solution for the CP phase $\xi (x)$ will be non-zero for some finite interval close to the origin. This signifies that the kink solution itself violates CP, even though the $Z_2$-symmetric 2HDM is CP invariant as well as it cannot realize spontaneous CP violation~\cite{TDLee, Branco:2011iw}.  Moreover, since $v_2(x)$ is an odd function of $x$ and $d\theta/dx$ is an even function
(due to the asymmetric boundaries),  one should expect that self-consistency of~\eqref{eq3.10} would require that $\xi (x)$ ($d\xi/dx$) is an odd (even) function
of $x$.

\begin{figure}
\centering
{\bf \subfloat[]{\includegraphics[width=0.5\textwidth]{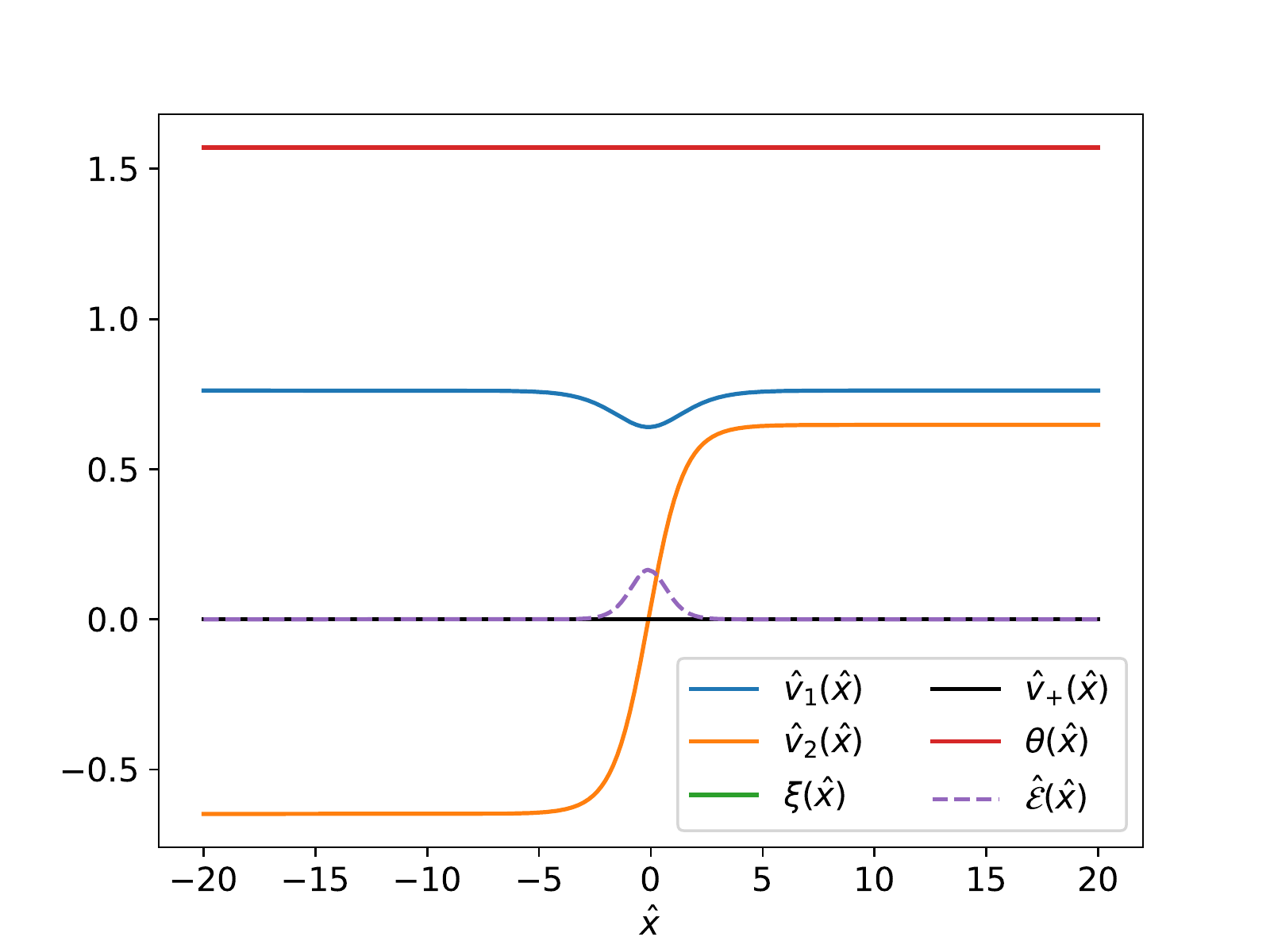}}
  \subfloat[]{\includegraphics[width=0.5\textwidth]{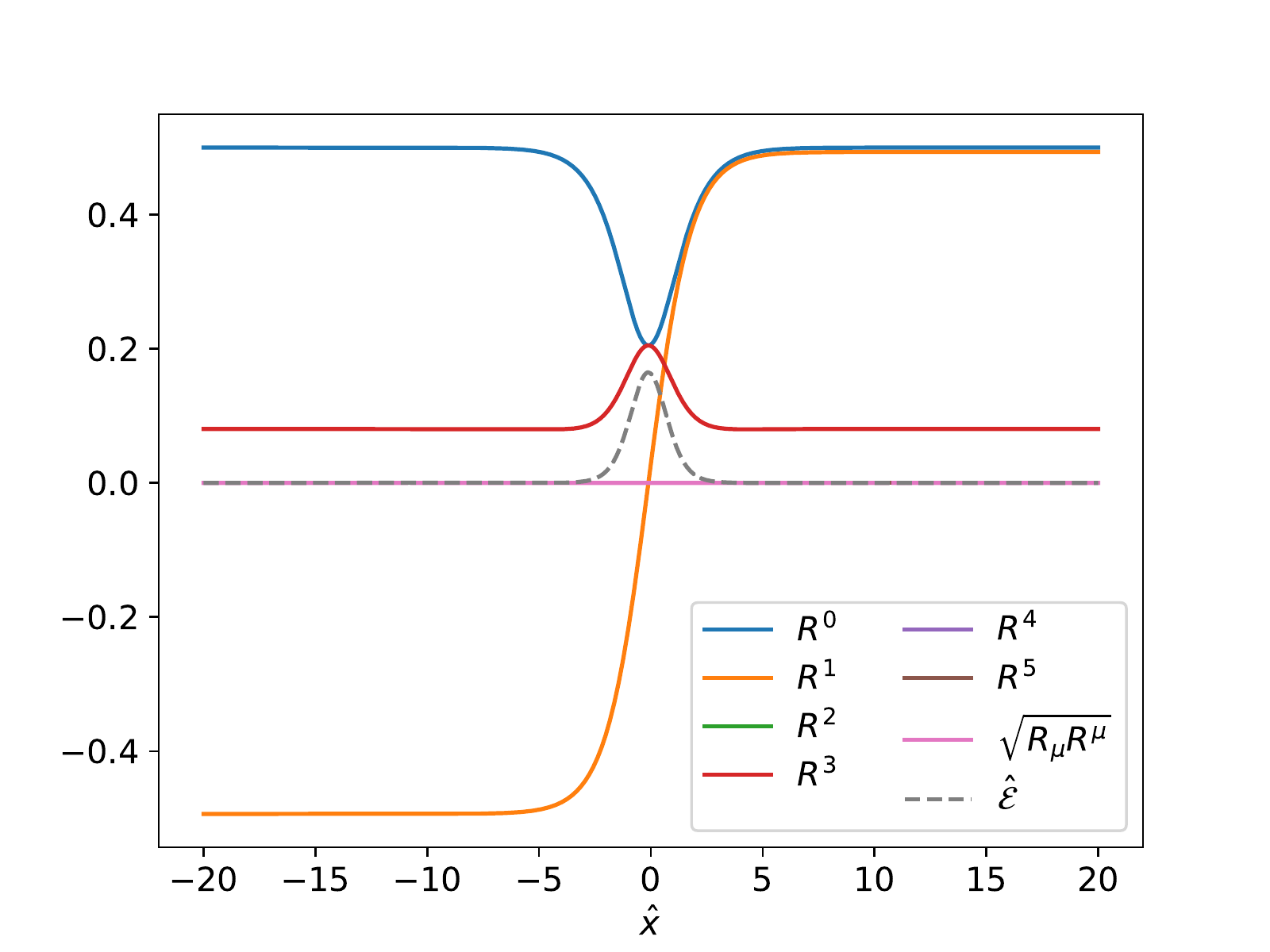}}
  \hfill
\subfloat[]{\includegraphics[width=0.5\textwidth]{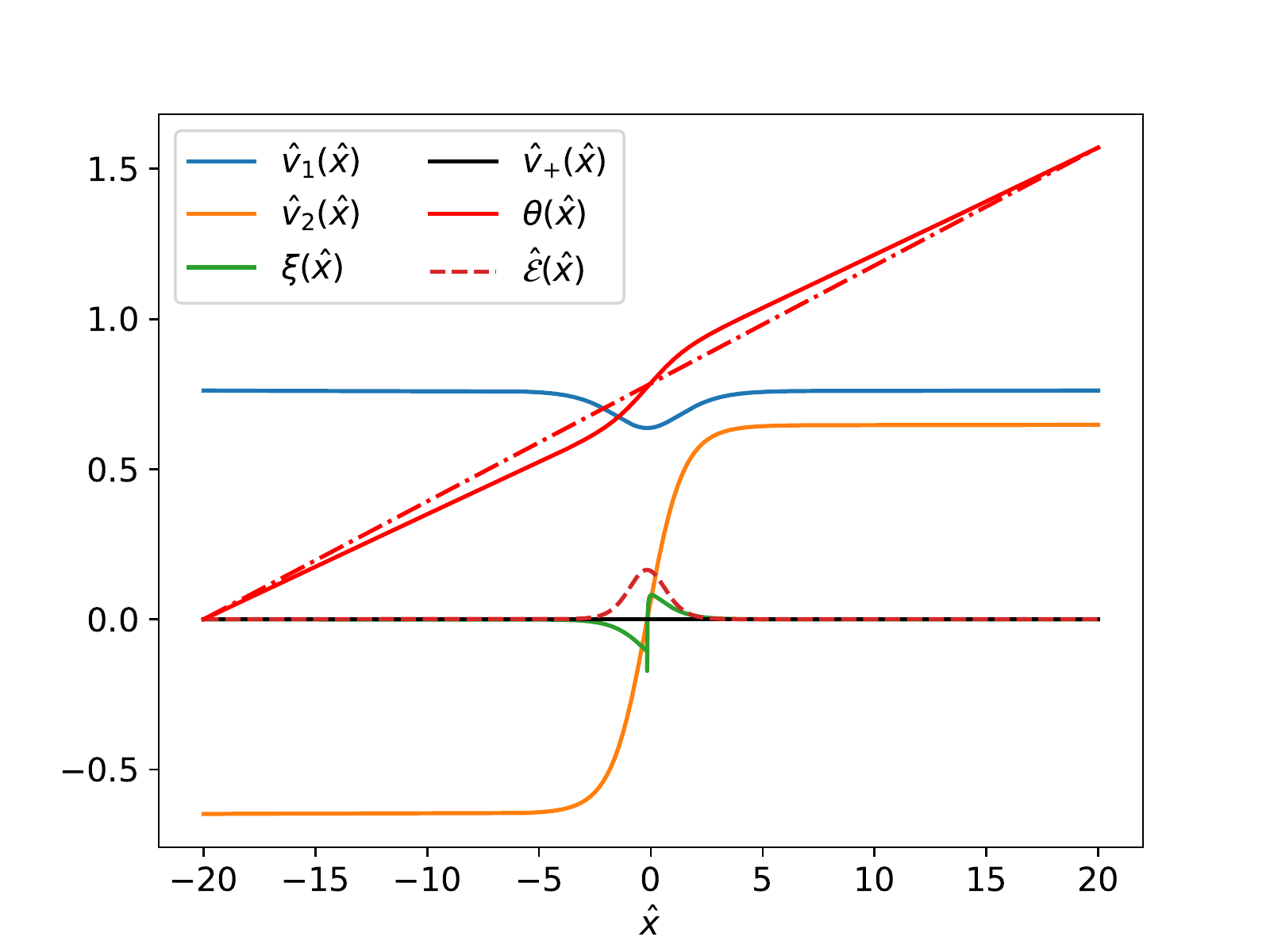}} 
\subfloat[]{\includegraphics[width=0.5\textwidth]{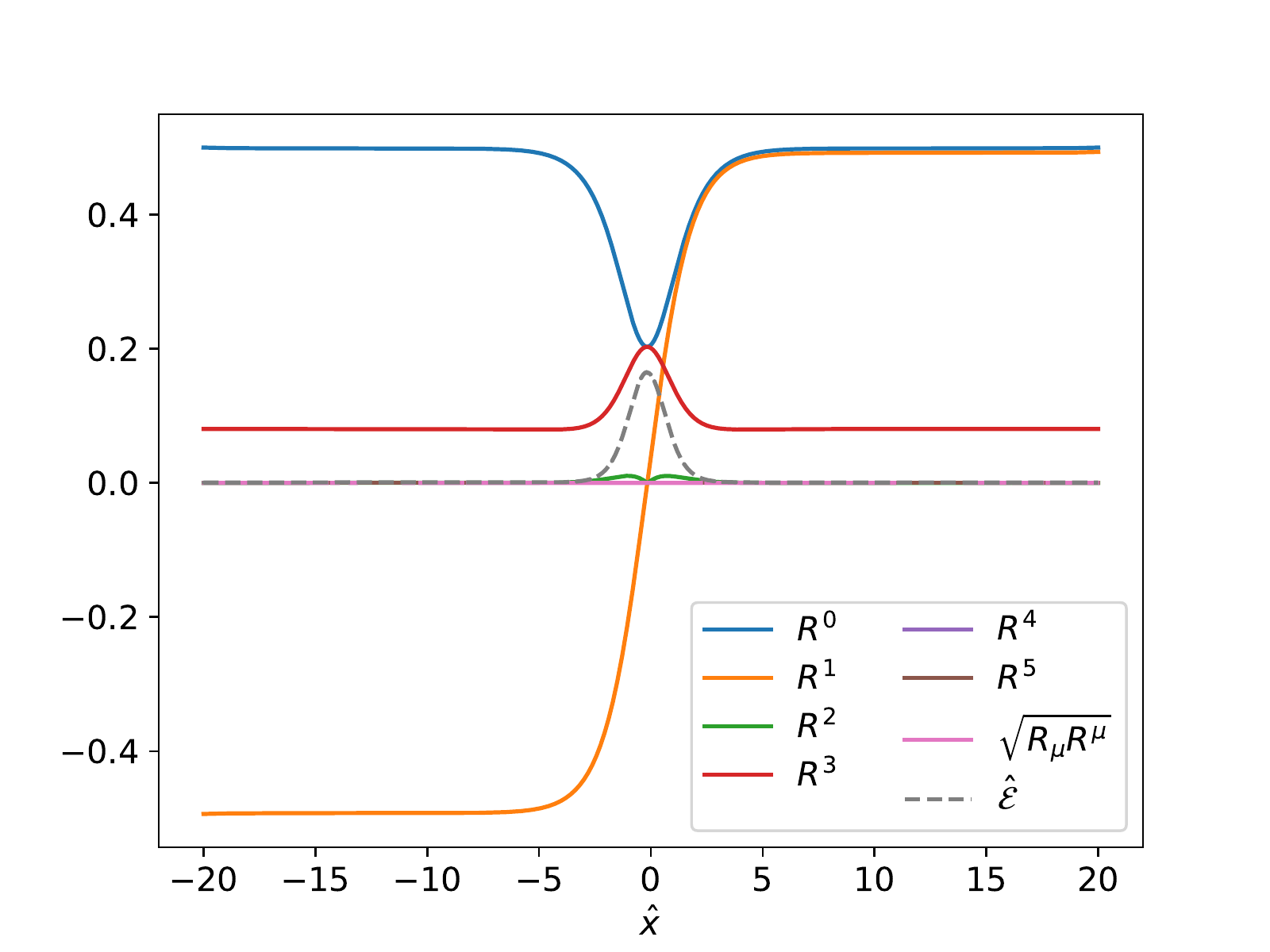}}} 
\caption{Numerical estimates of the vacuum parameters and the energy density using the gradient flow method for different boundary conditions on~$\theta (\hat{x})$: {\bf (a)}~$\theta(\hat{x})=\pi/2$ at both boundaries,  {\bf (b)}~$R$-field space profiles with $\theta(\hat{x})=\pi/2$ at both boundaries,  {\bf (c)}~$\theta(\hat{x})=0$ at the LH boundary and $\theta(\hat{x})=\pi/2$ at the RH boundary. The dash-dotted line through the boundary points is shown for comparison,  {\bf (d)}~$R$-field space profiles with $\theta(\hat{x})=0$ at the LH boundary and $\theta(\hat{x})=\pi/2$ at the RH boundary.}
\label{Figure6.1}
\end{figure} 

The above analytical observations may also be verified by our numerical simulations. To confirm these observations, we first impose symmetric boundary conditions, such that ${\theta(-L))=\theta(L) = \pi/2}$ at both the left hand (LH) and the right hand (RH) boundary of a finite interval $-L \le \hat{x} \le L$, with $\hat{x}\equiv M_h x$ (see Appendix~\ref{AppendixA}). Here, $L$ is a length cut-off in units of $M^{-1}_h$ that should be send to infinity upon completion of the simulation. The results of our analysis for the five vacuum parameters, $v_1, v_2, \xi, v_{+}$ and $\theta$, are shown in Figure \ref{Figure6.1}(a),
and the corresponding $x$-profiles in the bilinear $R$-space are displayed in
Figure~\ref{Figure6.1}(b). As expected, we find that the kink solutions in Figure~\ref{Figure6.1}(a) are both charge and CP-preserving, with $v_{+}(x) =0$ for all $x$. The latter is reflected in Figure~\ref{Figure6.1}(b), with the kink solution obeying the vacuum neutrality condition: $R_{\mu}R^{\mu}=0$.

Let us now impose an asymmetric boundary condition on $\theta (x)$, with $\theta (-L)=0$ at the LH boundary and $\theta(L) = \pi/2$ at the RH boundary. This corresponds to a relative U(1)$_Y$ gauge rotation of the vacua at infinity~\cite{SDW}. The initial guess function for $\theta(x,t=0)$ at the origin of time $t$ is chosen to be a straight line connecting the two boundaries at $x=\pm L$. The gradient flow numerical results are shown in Figure \ref{Figure6.1}(c), with the corresponding $R$-space profiles shown in Figure \ref{Figure6.1}(d).
As before, we see that the kink solution is electrically neutral with $v_{+}(x)=0$ for all $x$, satisfying the vacuum neutrality condition: $R_{\mu}R^{\mu}=0$. This should not be surprising, since $v_{+}(x)=0$ is the lowest energy configuration which is still compatible with all the imposed boundary conditions.

Nevertheless, we see from Figure~\ref{Figure6.1}(c) that $\xi(x)$ is a non-zero and odd function of $x$ for an interval close to the origin. This also gives rise to a non-zero $x$-profile in the same region for the component $R^{2}$ in the bilinear $R$-space. As a consequence, the so-determined kink solution is CP-violating, and its analytic behaviour agrees well with our discussion in connection with~\eqref{eq3.10}.  We also observe that the ground state solution for $\theta(\hat{x})$ tends asymptotically to a straight line with a non-zero slope $\pi/(4L)$ at the boundaries, $\hat{x} = \pm L$, where $L=20$ is the
length cut-off (in units of $M^{-1}_h$) used in our gradient flow analysis. This may cause some concern regarding the validity of our results. However, in realistic situations, we must send $L\rightarrow\infty$. Hence, we have \begin{equation} \label{eq:dthetaL}
  \frac{d\theta}{d\hat{x}}\bigg|_{\hat{x}=\pm L} =\,
  \frac{\pi}{4L}\: \rightarrow\: 0\,,
\end{equation}
as the length cut-off $L$ goes to infinity. In particular, given the analytic behaviour of
${d\theta/d\hat{x} \propto 1/L}$ in~\eqref{eq:dthetaL} at the boundaries, it is not difficult to check using~\eqref{eq11} that the total (kinetic) energy of the kink solution,
\begin{equation}
   \label{eq:totalE}
E_{\text{kin}}(L)\, =\, \int_{-L}^Ld\hat{x}\,\mathcal{E}_{\rm kin}(\hat{x})\, ,
\end{equation}
is proportional to $1/L$, and so remains finite in the limit $L\to \infty$, as it is expected
on general theoretical grounds.

\subsubsection{The $G^{1}$-Scenario}

Our second simplified scenario will be to consider the effect of a non-zero Goldstone
mode~$G^{1}(x)$, but take all other Goldstone modes to vanish, i.e.~by setting  $\theta (x), G^{2,3}(x) = 0$, for all $x$. In this $G^1$-scenario, the relevant electroweak gauge transformation matrix becomes
\begin{equation}\label{eq6.1.11}
  U =\exp\left(\frac{i\widehat{G}^{1}(x)\,\sigma^{1}}{2}\right)=\begin{pmatrix}
    \cos\left(\widehat{G}^{1}/2\right) & i\sin\left(\widehat{G}^{1}/2\right) \\ i\sin\left(\widehat{G}^{1}/2\right) & \cos\left(\widehat{G}^{1}/2\right) \end{pmatrix}\,.
\end{equation}
Substituting \eqref{eq6.1.11} into \eqref{eq11}, the kinetic energy density evaluates to
\begin{equation}
  \begin{split}
    \label{eq:EkinG1}
\mathcal{E}_{\textrm{kin}}=&\frac{1}{2}{\left(\frac{dv_{1}}{dx}\right)}^{2}+\frac{1}{2}{\left(\frac{dv_{2}}{dx}\right)}^{2}+\frac{1}{2}{\left(\frac{dv_{+}}{dx}\right)}^{2}+\frac{1}{2}v_{2}^{2}{\left(\frac{d\xi}{dx}\right)}^{2}+\frac{1}{8}(v_{1}^{2}+v_{2}^{2}+v_{+}^{2}){\left(\frac{d\widehat{G}^{1}}{dx}\right)}^{2}\\
&+\frac{1}{2}\left(v_{+}\sin\xi\frac{dv_{2}}{dx}+v_{+}v_{2}\cos\xi\frac{d\xi}{dx}-v_{2}\sin\xi\frac{dv_{+}}{dx}\right)\frac{d\widehat{G}^{1}}{dx}\; .
\end{split}
\end{equation}
For this second $G^1$-scenario, the gradient flow equations are found to be
\begin{equation}
\begin{split}
\frac{\partial v_1}{\partial t}\, =&\ \frac{\partial^{2}v_{1}}{\partial x^{2}}-\frac{1}{4}v_{1}{\left(\frac{\partial\widehat{G}^{1}}{\partial x}\right)}^{2}+\mu_{1}^{2}v_{1}-\lambda_{1}v_{1}^{3}-\frac{1}{2}\lambda_{3}v_{1}v_{+}^{2}-\frac{1}{2}(\lambda_{34}-|\lambda_{5}|c_{2\xi})v_{1}v_{2}^{2}\,,\\
\frac{\partial v_{2}}{\partial t}\, =&\ \frac{\partial^{2}v_{2}}{\partial x^{2}}+\frac{1}{2}v_{+}\sin\xi\frac{\partial^{2}\widehat{G}^{1}}{\partial x^{2}}-v_{2}{\left(\frac{\partial\xi}{\partial x}\right)}^{2}-\frac{1}{4}v_{2}{\left(\frac{\partial\widehat{G}^{1}}{\partial x}\right)}^{2}+\sin\xi\frac{\partial\widehat{G}^{1}}{\partial x}\frac{\partial v_{+}}{\partial x}\\
&+\mu_{2}^{2}v_{2}-\lambda_{2}v_{2}(v_{2}^{2}+v_{+}^{2})
-\frac{1}{2}(\lambda_{34}-|\lambda_{5}|c_{2\xi})v_{1}^{2}v_{2}\,,\\ 
\frac{\partial v_{+}}{\partial t}\, =&\ \frac{\partial^{2}v_{+}}{\partial x^{2}}-\frac{1}{2}v_{2}\sin\xi\frac{\partial^{2}\widehat{G}^{1}}{\partial x^{2}}-\frac{1}{4}v_{+}{\left(\frac{\partial\widehat{G}^{1}}{\partial x}\right)}^{2}-\sin\xi\frac{\partial\widehat{G}^{1}}{\partial x}\frac{\partial v_{2}}{\partial x}-v_{2}\cos\xi\frac{\partial\widehat{G}^{1}}{\partial x}\frac{\partial\xi}{\partial x}\\
&+\mu_{2}^{2}v_{+}-\lambda_{2}v_{+}(v_{2}^{2}+v_{+}^{2})-\frac{1}{2}\lambda_{3}v_{1}^{2}v_{+}\,,\\
\frac{\partial\xi}{\partial t}\, =&\ \ v_{2}^{2}\frac{\partial^{2}\xi}{\partial x^{2}}+2v_{2}\frac{\partial v_{2}}{\partial x}\frac{\partial\xi}{\partial x}+\frac{1}{2}v_{+}v_{2}\cos\xi\frac{\partial^{2}\widehat{G}^{1}}{\partial x^{2}}+v_{2}\cos\xi\frac{\partial v_{+}}{\partial x}\frac{\partial\widehat{G}^{1}}{\partial x}-\frac{1}{2}|\lambda_{5}|v_{1}^{2}v_{2}^{2}s_{2\xi}\,,\\[3mm]
\frac{\partial\widehat{G}^{1}}{\partial t}\, =&\ \frac{1}{4}(v_{1}^{2}+v_{2}^{2}+v_{+}^{2})\frac{\partial^{2}\widehat{G}^{1}}{\partial x^{2}}+\frac{\partial}{\partial x}\left(\frac{1}{2}v_{+}\sin\xi\frac{\partial v_{2}}{\partial x}+\frac{1}{2}v_{+}v_{2}\cos\xi\frac{\partial\xi}{\partial x}-\frac{1}{2}v_{2}\sin\xi\frac{\partial v_{+}}{\partial x}\right)\\
&+\frac{1}{2}\left(v_{1}\frac{\partial v_{1}}{\partial x}+v_{2}\frac{\partial v_{2}}{\partial x}+v_{+}\frac{\partial v_{+}}{\partial x}\right)\left(\frac{\partial\widehat{G}^{1}}{\partial x}\right)\, .
\end{split}
\end{equation}

In order to obtain a finite-energy kink solution, the derivatives of all Goldstone bosons and vacuum parameters should tend to zero at the boundaries. In the ground state, the gradient flow equation for $\widehat{G}^{1}$ takes the form:
\begin{equation}
\frac{1}{4}(v_{1}^{2}+v_{2}^{2}+v_{+}^{2})\,\frac{d\widehat{G}^{1}}{dx}\: -\: \frac{1}{2}v_{2}^{2}\sin^{2}\xi\frac{d}{d x}\left(\frac{v_{+}}{v_{2}\sin\xi}\right)\: =\: 0\,.
\end{equation}
We may now solve this last equation for $d\widehat{G}^{1}/dx$,
\begin{equation}\label{eq46}
  \frac{d\widehat{G}^{1}}{dx}\ =\ \frac{2v_{2}^{2}\sin^{2}\xi}{v_{1}^{2}+v_{2}^{2}+v_{+}^{2}}\,
  \frac{d}{dx}\left(\frac{v_{+}}{v_{2}\sin\xi}\right)\,.
\end{equation}
To get the lowest contribution from~$\xi(x)$ to $\mathcal{E}_{\textrm{kin}}$ in~\eqref{eq:EkinG1} compatible with its vanishing boundary conditions at $x = \pm L$, one must have $d\xi (x)/dx = \xi(x) = 0$ for all $x$. If $d\widehat{G}^{1}/dx$ varies significantly for some finite interval of $x$ enforced by the asymmetric boundary values of $G^1(x)$ at $x = \pm L$, then the analytic property~\eqref{eq46} can only hold true when $v_+ (x)$ does not vanish everywhere in~$x$. In this asymmetric $G^1$-scenario, one may expect that the kink solution violates charge conservation, but respects CP. 

\begin{figure}
{\bf \subfloat[]{\includegraphics[width=0.5\textwidth]{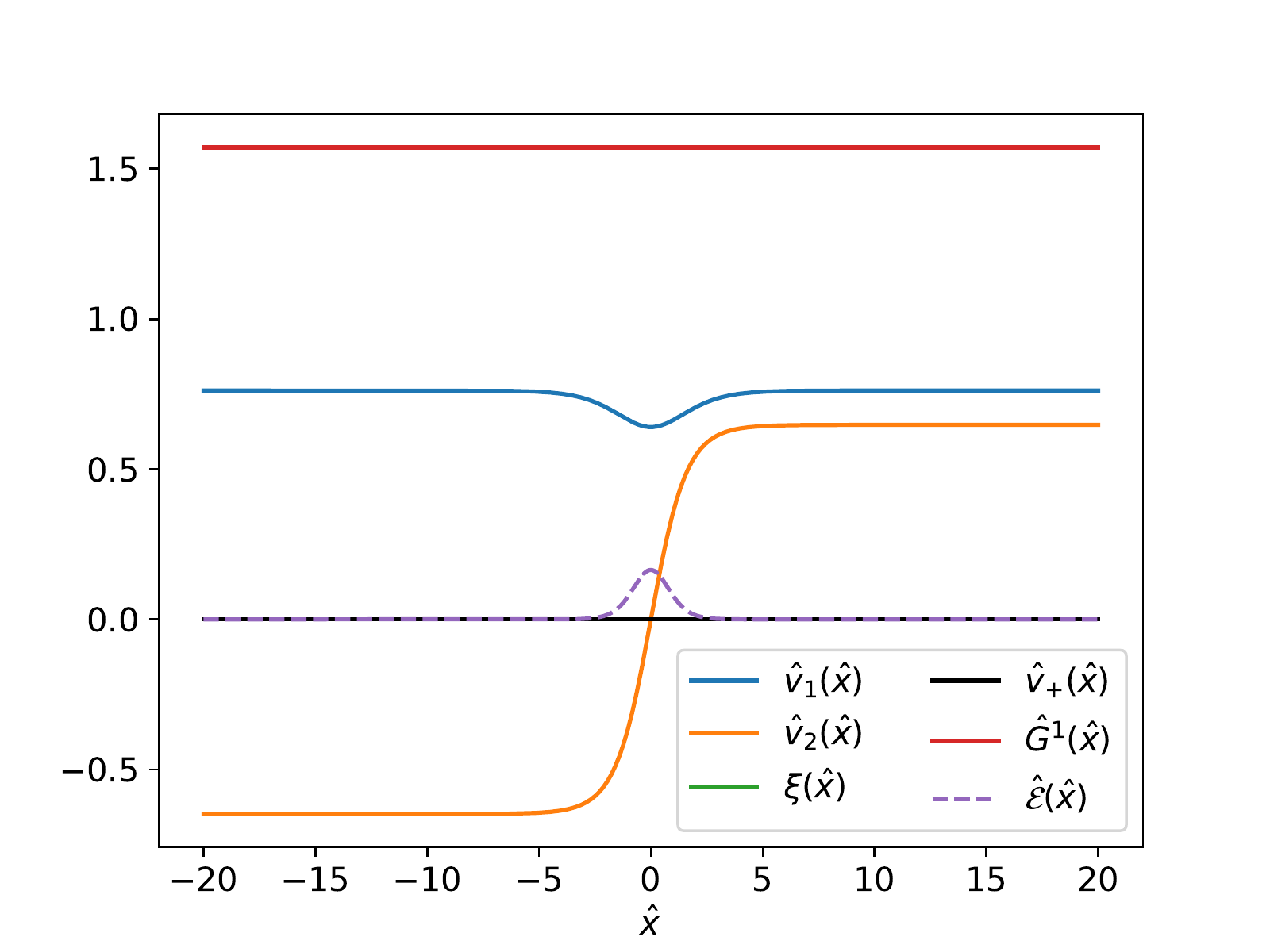}}
\subfloat[]{\includegraphics[width=0.5\textwidth]{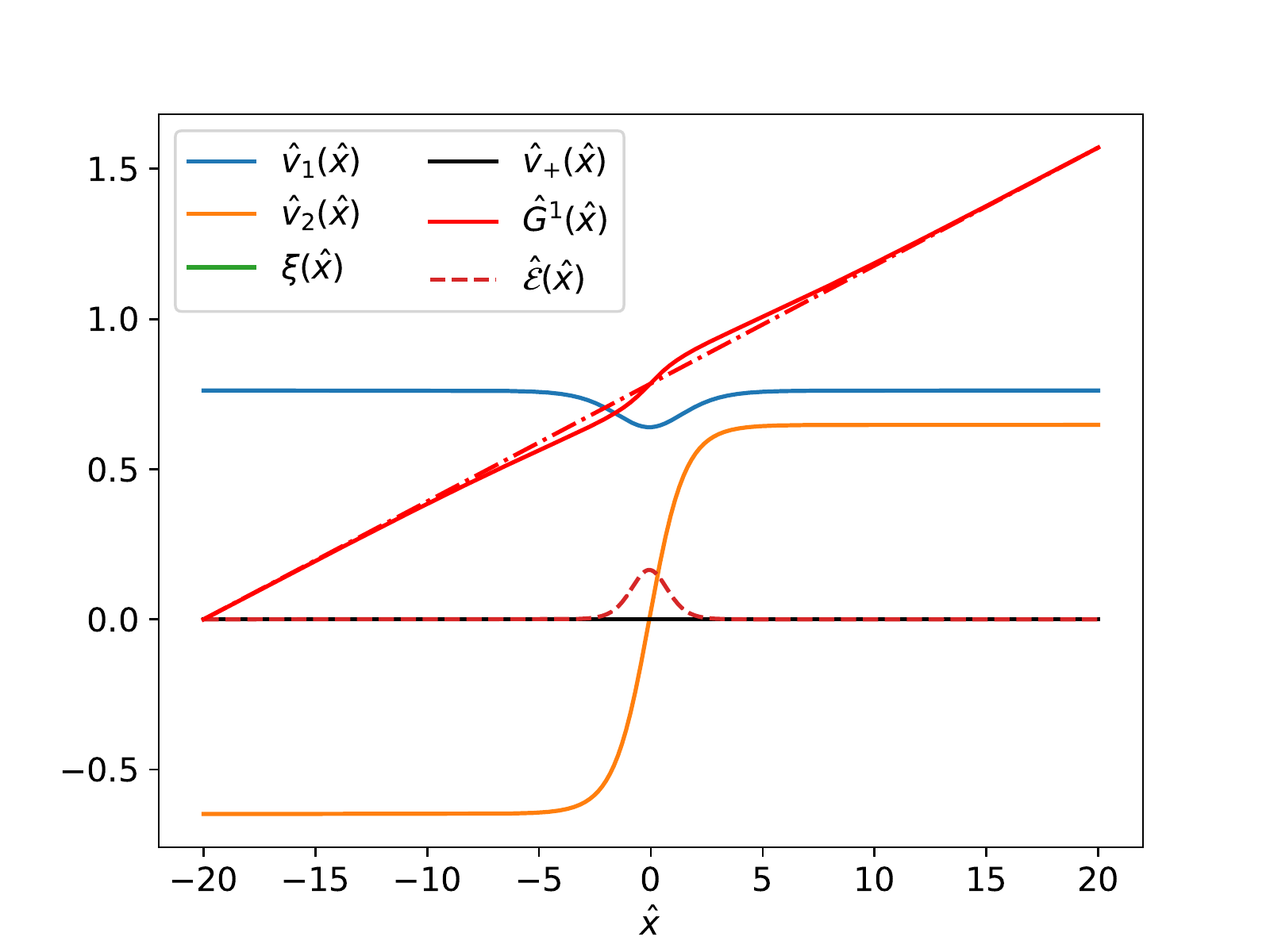}} }
\caption{Numerical estimates of the kink parameters for two different boundary
  conditions on~$\widehat{G}^{1}(\hat{x})$: {\bf (a)}~$\widehat{G}^{1}(\hat{x})=\pi/2$ at both boundaries, {\bf (b)}~$\widehat{G}^{1}(\hat{x})=0$ at the LH boundary and ${\widehat{G}^{1}(\hat{x})=\pi/2}$ at the RH boundary. The dash-dotted line
  through the two boundary points is also shown for comparison.}
\label{Figure6.2}
\end{figure}

The results of a gradient flow analysis are shown in Figure~\ref{Figure6.2}.  Figure~\ref{Figure6.2}(a) presents a numerical simulation using symmetric boundary conditions $\widehat{G}^{1}=\pi/2$ at both LH and RH boundaries. Instead, Figure~\ref{Figure6.2}(c) displays a simulation by imposing asymmetric boundary conditions, with~$\widehat{G}^{1}=0$ at the LH boundary and $\widehat{G}^{1}=\pi/2$ at the RH boundary. No charge violation was noticeable for both types of symmetric and asymmetric boundary conditions on $\widehat{G}^{1}$, i.e.~$v_+(x) = 0$ for all $x$.  In addition, the CP-odd vacuum parameter $\xi (x)$ is also zero everywhere in $x$, according to our discussion given above. However, for the asymmetric case, the finding of an unobservable charge violation is a bit unexpected, but it may be attributed to a good extent to the dispersive (non-localised) feature of the obtained solution for $\widehat{G}^{1}(x)$. The $G^1(x)$ solution from a gradient flow computation has a non-vanishing and almost constant slope $\pi/(4L)$, for the entire $x$-interval $(-L\,,L)$. Hence, $dG^1/dx$ will vanish for all $x$, as $L\to \infty$. Moreover, the kinetic and total energies of the kink can then easily be shown to remain finite in the same limit for $L$.

\subsubsection{The $G^{2}$-Scenario}

In our third $G^2$-scenario, we study the effect of $G^2(x)$ alone by setting $\theta (x), G^{1,3}(x) = 0$, for all~$x$. In this case, the unitary matrix $U$ describing electroweak gauge transformations assumes the SO(2) form:
\begin{equation}
  \label{eq:UforG2}
  U\,=\,\exp\left(\frac{i\widehat{G}^{2}(x) \sigma^{2}}{2}\right)\,=\,
  \begin{pmatrix}\cos\left(\widehat{G}^{2}/2\right) & \sin\left(\widehat{G}^{2}/2\right) \\ -\sin\left(\widehat{G}^{2}/2\right) & \cos\left(\widehat{G}^{2}/2\right) \end{pmatrix}\,.
\end{equation}
Taking this last expression of $U$ into account, the kinetic energy density becomes
\begin{equation}
    \label{eq6.1.19}
  \begin{split}
    \mathcal{E}_{\textrm{kin}}\: =\:  &\frac{1}{2}{\left(\frac{dv_{1}}{dx}\right)}^{2}+\frac{1}{2}{\left(\frac{dv_{2}}{dx}\right)}^{2}+\frac{1}{2}{\left(\frac{dv_{+}}{dx}\right)}^{2}+\frac{1}{2}v_{2}^{2}{\left(\frac{d\xi}{dx}\right)}^{2}+\frac{1}{8}(v_{1}^{2}+v_{2}^{2}+v_{+}^{2}){\left(\frac{d\widehat{G}^{2}}{dx}\right)}^{2}\\
    &+\frac{1}{2}\left(v_{2}\cos\xi\frac{dv_{+}}{dx}-v_{+}\cos\xi\frac{dv_2}{dx}+v_{2}v_{+}\sin\xi\frac{d\xi}{dx}\right)\frac{d\widehat{G}^{2}}{dx}\;. \end{split}
\end{equation}
For this $G^2$-scenario, the gradient flow equations may be cast into the form:
\begin{equation}
\begin{split}
\frac{\partial v_1}{\partial t}\, =&\ \frac{\partial^{2}v_{1}}{\partial x^{2}}-\frac{1}{4}v_{1}{\left(\frac{\partial\widehat{G}^{2}}{\partial x}\right)}^{2}+\mu_{1}^{2}v_{1}-\lambda_{1}v_{1}^{3}-\frac{1}{2}\lambda_{3}v_{1}v_{+}^{2}-\frac{1}{2}(\lambda_{34}-|\lambda_{5}|c_{2\xi})v_{1}v_{2}^{2}\,,\\
\frac{\partial v_{2}}{\partial t}\, =&\ \frac{\partial^{2}v_{2}}{\partial x^{2}}-v_{2}{\left(\frac{\partial\xi}{\partial x}\right)}^{2}-\frac{1}{2}v_{+}\cos\xi\frac{\partial^{2}\widehat{G}^{2}}{\partial x^{2}}-\frac{1}{4}v_{2}{\left(\frac{\partial\widehat{G}^{2}}{\partial x}\right)}^{2}-\cos\xi\frac{\partial v_{+}}{\partial x}\frac{\partial\widehat{G}^{2}}{\partial x} \\
&+\mu_{2}^{2}v_{2}-\lambda_{2}v_{2}(v_{2}^{2}+v_{+}^{2})-\frac{1}{2}(\lambda_{34}-|\lambda_{5}|c_{2\xi})v_{1}^{2}v_{2}\,,\\
\frac{\partial v_{+}}{\partial t}\,=&\ \frac{\partial^{2}v_{+}}{\partial x^{2}}+\frac{1}{2}v_{2}\cos\xi\frac{\partial^{2}\widehat{G}^{2}}{\partial x^{2}}-\frac{1}{4}v_{+}{\left(\frac{\partial\widehat{G}^{2}}{\partial x}\right)}^{2}+\cos\xi\frac{\partial v_{2}}{\partial x}\frac{\partial\widehat{G}^{2}}{\partial x}-v_{2}\sin\xi\frac{\partial\xi}{\partial x}\frac{\partial\widehat{G}^{2}}{\partial x} \\
&+\mu_{2}^{2}v_{+}-\lambda_{2}v_{+}(v_{2}^{2}+v_{+}^{2})-\frac{1}{2}\lambda_{3}v_{1}^{2}v_{+}\,,\\
\frac{\partial\xi}{\partial t}\, =&\ v_{2}^{2}\frac{\partial^{2}\xi}{\partial x^{2}}+\frac{1}{2}v_{2}v_{+}\sin\xi\frac{\partial^{2}\widehat{G}^{2}}{\partial x^{2}}+2v_{2}\frac{\partial v_{2}}{\partial x}\frac{\partial\xi}{\partial x}+v_{2}\sin\xi\frac{\partial v_{+}}{\partial x}\frac{\partial\widehat{G}^{2}}{\partial x}-\frac{1}{2}|\lambda_{5}|v_{1}^{2}v_{2}^{2}s_{2\xi}\,,\\[3mm]
\frac{\partial\widehat{G}^{2}}{\partial t}\, =&\ \frac{1}{4}(v_{1}^{2}+v_{2}^{2}+v_{+}^{2})\frac{\partial^{2}\widehat{G}^{2}}{\partial x^{2}}+\frac{1}{2}\left(v_{1}\frac{\partial v_{1}}{\partial x}+v_{2}\frac{\partial v_{2}}{\partial x}+v_{+}\frac{\partial v_{+}}{\partial x}\right)\left(\frac{\partial\widehat{G}^{2}}{\partial x}\right)\\
&+\frac{1}{2}\frac{\partial}{\partial x}\left(v_{2}\cos\xi\frac{\partial v_{+}}{\partial x}-v_{+}\cos\xi\frac{\partial v_2}{\partial x}+
  v_{2}v_{+}\sin\xi\frac{\partial\xi}{\partial x}\right)\,.
\end{split}
\end{equation}

Exactly as we did before, the following constraining equation for~$\widehat{G}^{2}(x)$
may be derived in the ground state:
\begin{equation}\label{eq43}
\frac{d\widehat{G}^{2}}{dx}\: =\: -\frac{2v_{2}^{2}\cos^{2}\xi}{v_{1}^{2}+v_{2}^{2}+v_{+}^{2}}\,\frac{d}{d x}\left(\frac{v_{+}}{v_{2}\cos\xi}\right)\;.
\end{equation}
Following a line of argumentation as we did before for the $G^1$-scenario,
Equation~\eqref{eq43} implies that the kink solution should respect CP,
but it can violate charge conservation, only when asymmetric boundary conditions
are chosen {\em and} $d\widehat{G}^{2}/dx \neq 0$ for a localised finite interval of $x$
as $L\to \infty$.

\begin{figure}
{\bf \subfloat[]{\includegraphics[width=0.5\textwidth]{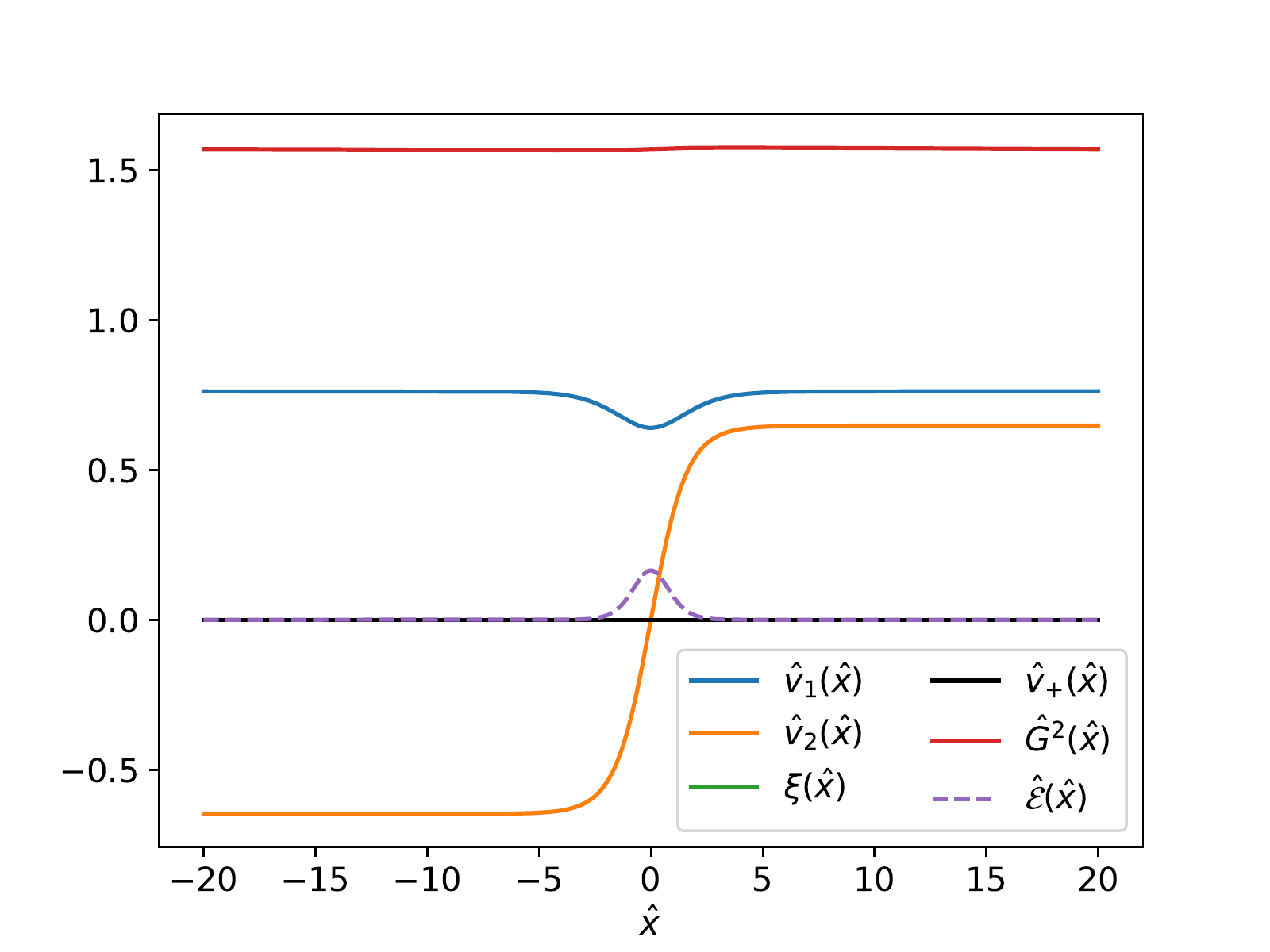}}
\subfloat[]{\includegraphics[width=0.5\textwidth]{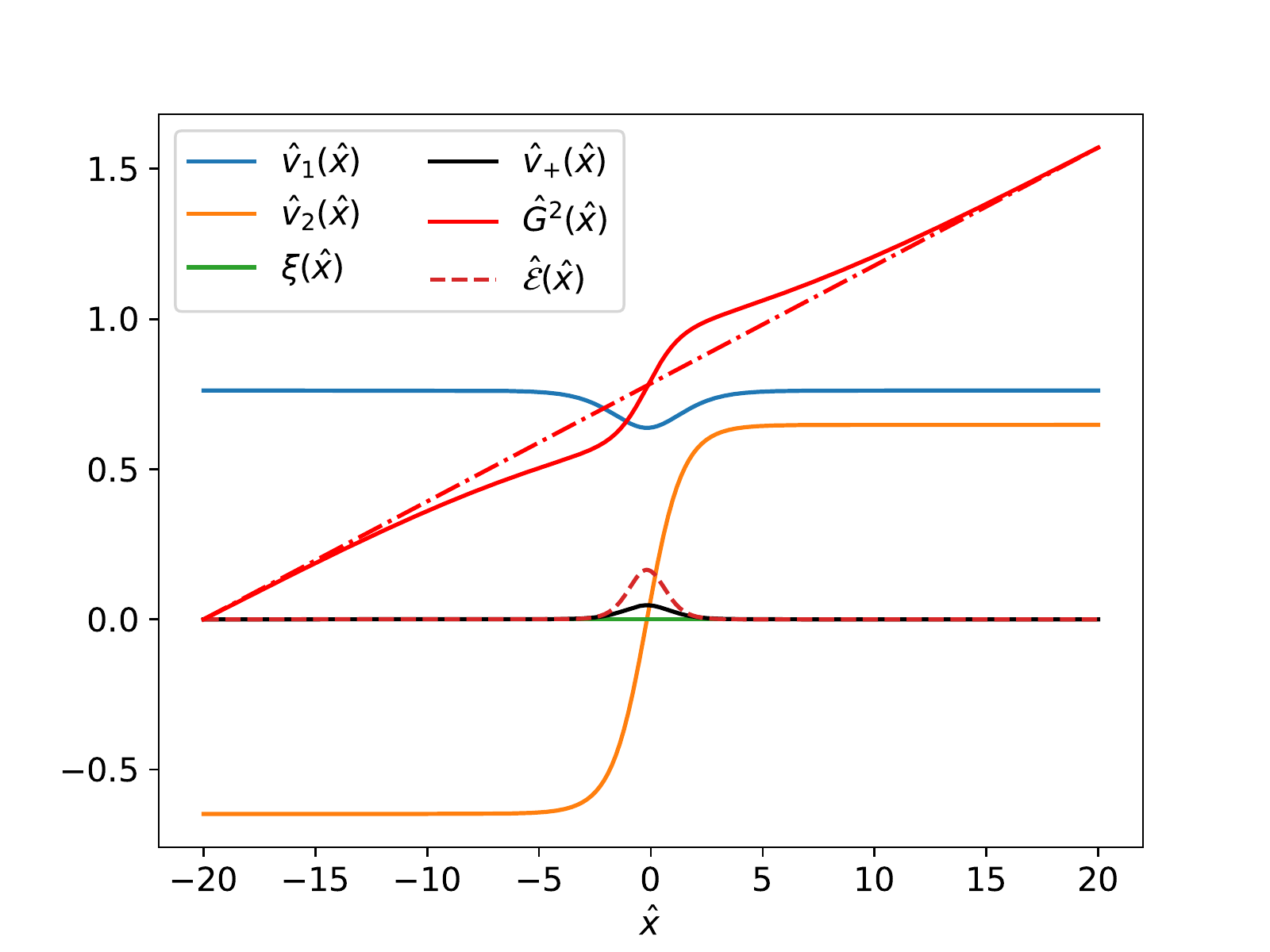}}
\hfill
\subfloat[]{\includegraphics[width=0.5\textwidth]{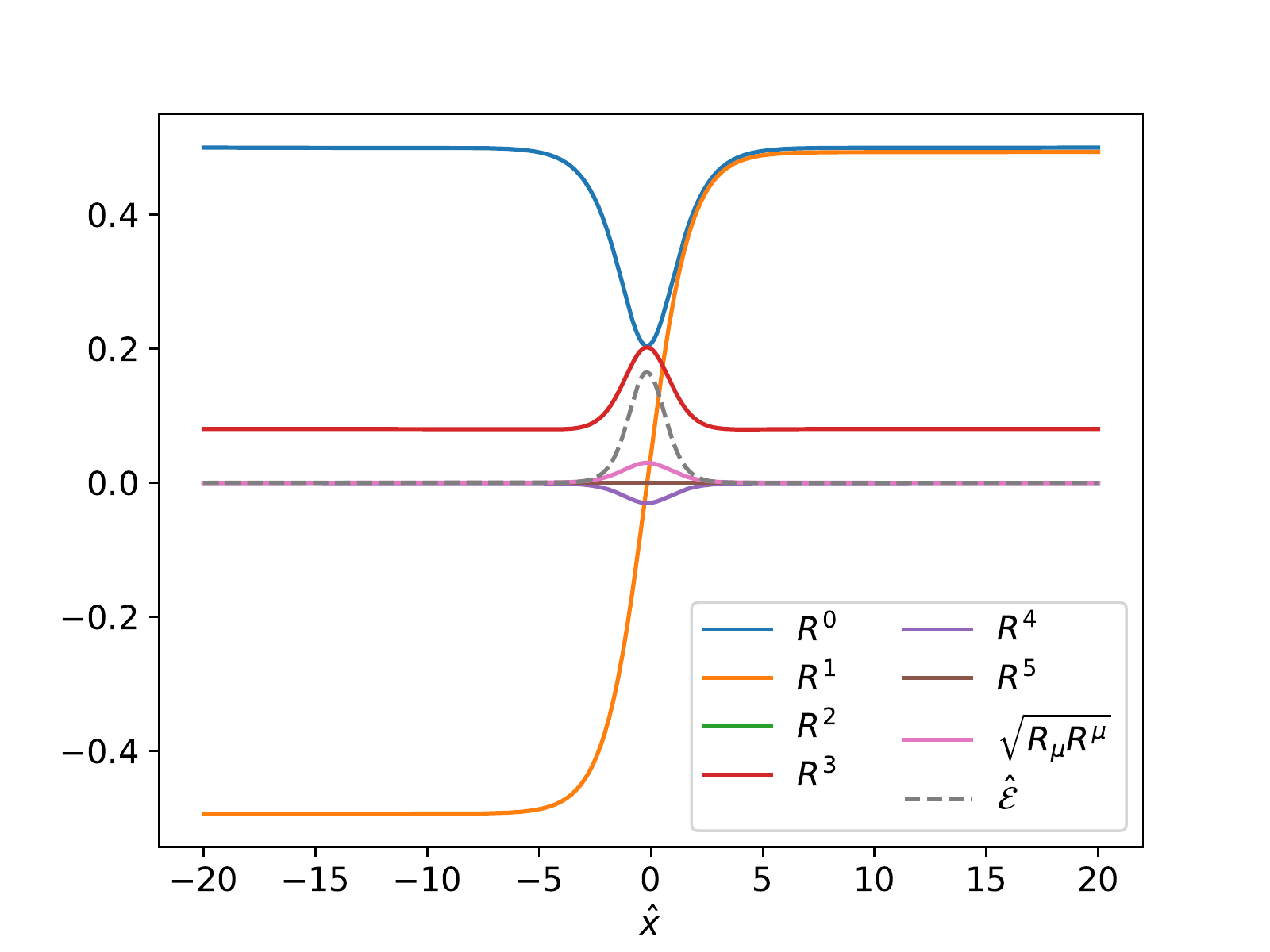}}
\subfloat[]{\includegraphics[width=0.5\textwidth]{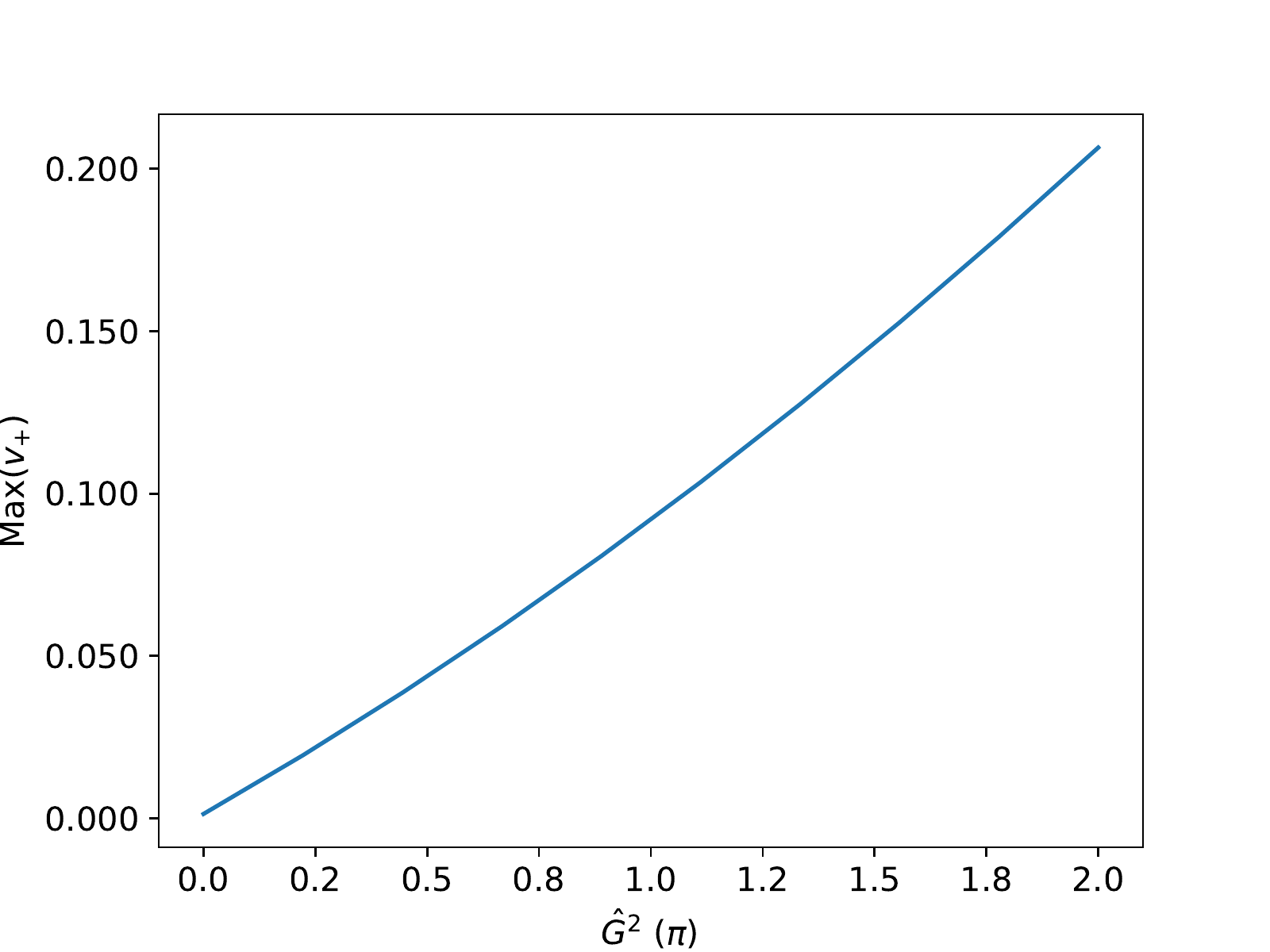}} }
\caption{Numerical estimates of the kink parameters for different boundary conditions on $\widehat{G}^{2}(\hat{x})$: {\bf (a)}~$\widehat{G}^{2}(\hat{x})=\frac{\pi}{2}$ at both boundaries, {\bf (b)}~$\widehat{G}^{2}(\hat{x})=0$ at the LH boundary and $\widehat{G}^{2}(\hat{x})
  = \pi /2$ at the RH boundary, {\bf (c)}~$R$-field space profiles with $\widehat{G}^{2}(\hat{x})=0$ at the LH boundary and $\widehat{G}^{2}(\hat{x})=\pi /2$ at the RH boundary, {\bf (d)}~maximum value for $v_{+}$ as a function of the value of $\widehat{G}^{2}$ at the RH boundary.}
\label{Figure6.3}
\end{figure}

We investigate two different cases by imposing symmetric and asymmetric boundary values
on $\widehat{G}^{2}(x)$. The $G^2$ solution preserves the neutrality of the ground state,
i.e.~$v_+(x) = 0$ for all $x$, for the symmetric case when $\widehat{G}^{2}$ is set to $\pi/2$ at both boundaries, as can be seen from Figure~\ref{Figure6.3}(a). However, when asymmetric gauge rotated vacua are selected at the boundaries with $\widehat{G}^2(-L) = 0$ and $\widehat{G}^2(+L) = \pi/2$, we then observe in Figure~\ref{Figure6.3}(b) a localised violation of charge, i.e.~$v_+(x) \neq 0$ close to the origin, having the same width as
the $v_2(x)$ kink solution. Instead, the CP phase $\xi (x) = 0$ everywhere respecting
CP invariance. These analytic properties are consistent with those derived from~\eqref{eq43}. 

We should comment here that our results shown in Figure~\ref{Figure6.3} are in good agreement with those presented in~\cite{SDW} (see, e.g., Figure 10), where a peak in $v_{+}$ was observed when a relative gauge rotation of~$\pi/2$ was applied to $\gamma_{1}$ at the RH boundary. Note that the angle $\gamma_{1}$ in~\cite{SDW} corresponds to~$\widehat{G}^{2}/2$ here, as long as all the other electroweak group parameters are set to zero, i.e.~$\gamma_{2}=\gamma_{3}=\theta=0$.
We complement the analysis given in~\cite{SDW} by showing explicitly the $x$-profile of
$\widehat{G}^{2}(x)$, so to better assess its direct impact on $v_+(x)$, by virtue of~\eqref{eq43}.

The lowest total energy $E_{\rm tot}$ of the 2HDM kink solution is obtained for $v_+=0$, in which case we have $E_{\rm tot} = 0.354$ (in $M_h$ units) per unit area. As shown in~Figure \ref{Figure6.3}(d), the maximum value of $v_+$ increases for large asymmetric boundary values of $G^2$.  For this asymmetric $G^2$-scenario, we have $\sqrt{R_{\mu}R^{\mu}}=|R^{4}|$, since both $\xi(x)$ and $\theta(x)$ are zero for all~$x$, which is a relation satisfied by our $R$-field space profiles shown in Figure~\ref{Figure6.3}(c).

\subsubsection{The $G^{3}$-Scenario}

Our last scenario of interest to us is the case where only the Goldstone field $G^3(x)$ is
non-zero, whereas all other would-be Goldstone modes vanish, i.e.~$\theta(x),\, \widehat{G}^{1,2}(x) =0$. For~this $G^3$-scenario, the electroweak gauge transformation matrix $U$ is
given by
\begin{equation}
   \label{eq:UforG3}
U\, =\, \exp\left(\frac{i\widehat{G}^{3}(x)\sigma^{3}}{2}\right)\: =\:  \begin{pmatrix}e^{i\widehat{G}^3/2} & 0 \\ 0 & e^{-i\widehat{G}^3/2} \end{pmatrix}\,.
\end{equation}
In this case, the kinetic energy density becomes
\begin{equation}
    \label{eq:EkinG3}
\begin{split}
\mathcal{E}_{\textrm{kin}}\, =&\, \frac{1}{2}{\left(\frac{dv_{1}}{dx}\right)}^{2}+\frac{1}{2}{\left(\frac{dv_{2}}{dx}\right)}^{2}+\frac{1}{2}{\left(\frac{dv_{+}}{dx}\right)}^{2}+\frac{1}{2}v_{2}^{2}{\left(\frac{d\xi}{dx}\right)}^{2}+\frac{1}{8}(v_{1}^{2}+v_{2}^{2}+v_{+}^{2}){\left(\frac{d\widehat{G}^{3}}{dx}\right)}^{2}\\  
& -\frac{1}{2}v_{2}^{2}\frac{d\xi}{dx}\frac{d\widehat{G}^{3}}{dx}\; .
\end{split}
\end{equation}
Given the form of $U$ in~\eqref{eq:UforG3}, we may derive a new set of gradient flow equations,
\begin{equation}
\begin{split}
\frac{\partial v_{1}}{\partial t}\, =&\ \frac{\partial^{2}v_{1}}{\partial x^{2}}-\frac{1}{4}v_{1}{\left(\frac{\partial\widehat{G}^{3}}{\partial x}\right)}^{2}+\mu_{1}^{2}v_{1}-\lambda_{1}v_{1}^{3}-\frac{1}{2}\lambda_{3}v_{1}v_{+}^{2}-\frac{1}{2}(\lambda_{34}-|\lambda_{5}|c_{2\xi})v_{1}v_{2}^{2}\,, \\
\frac{\partial v_{2}}{\partial t}\, =&\ \frac{\partial^{2}v_{2}}{\partial x^{2}}-v_{2}{\left(\frac{\partial\xi}{\partial x}\right)}^{2}-\frac{1}{4}v_{2}{\left(\frac{d\widehat{G}^{3}}{\partial x}\right)}^{2}+v_{2}\frac{\partial\xi}{\partial x}\frac{\partial\widehat{G}^{3}}{\partial x}+\mu_{2}^{2}v_{2}-\lambda_{2}v_{2}(v_{2}^{2}+v_{+}^{2})\\
&-\frac{1}{2}(\lambda_{34}-|\lambda_{5}|c_{2\xi})v_{1}^{2}v_{2}\,,\\
\frac{\partial v_{+}}{\partial t}\, =&\ \frac{\partial^{2}v_{+}}{\partial x^{2}}-\frac{1}{4}v_{+}{\left(\frac{\partial\widehat{G}^{3}}{\partial x}\right)}^{2}+\mu_{2}^{2}v_{+}-\lambda_{2}v_{+}(v_{2}^{2}+v_{+}^{2})
-\frac{1}{2}\lambda_{3}v_{1}^{2}v_{+}\,,\\ 
\frac{\partial\xi}{\partial t}\, =&\ v_{2}^{2}\frac{\partial^{2}\xi}{\partial x^{2}}-\frac{1}{2}v_{2}^{2}\frac{\partial^{2}\widehat{G}^{3}}{\partial x^{2}}+2v_{2}\frac{\partial v_{2}}{\partial x}\frac{\partial\xi}{\partial x}-v_{2}\frac{\partial v_{2}}{\partial x}\frac{\partial\widehat{G}^{3}}{\partial x}-\frac{1}{2}|\lambda_{5}|v_{1}^{2}v_{2}^{2}s_{2\xi}\,,\\[3mm] 
\frac{\partial\widehat{G}^{3}}{\partial t} \, =&\ \frac{1}{4}\left(v_{1}^{2}+v_{2}^{2}+v_{+}^{2}\right)\frac{\partial^{2}\widehat{G}^{3}}{\partial x^{2}}-\frac{1}{2}v_{2}^{2}\frac{\partial^{2}\xi}{\partial x^{2}}+\frac{1}{2}\left(v_{1}\frac{\partial v_{1}}{\partial x}+v_{2}\frac{\partial v_{2}}{\partial x}+v_{+}\frac{\partial v_{+}}{\partial x}\right)\frac{\partial\widehat{G}^{3}}{\partial x}-v_{2}\frac{\partial v_{2}}{\partial x}\frac{\partial\xi}{\partial x}\, .
\end{split}
\end{equation}

\begin{figure}
{\bf \subfloat[]{\includegraphics[width=0.5\textwidth]{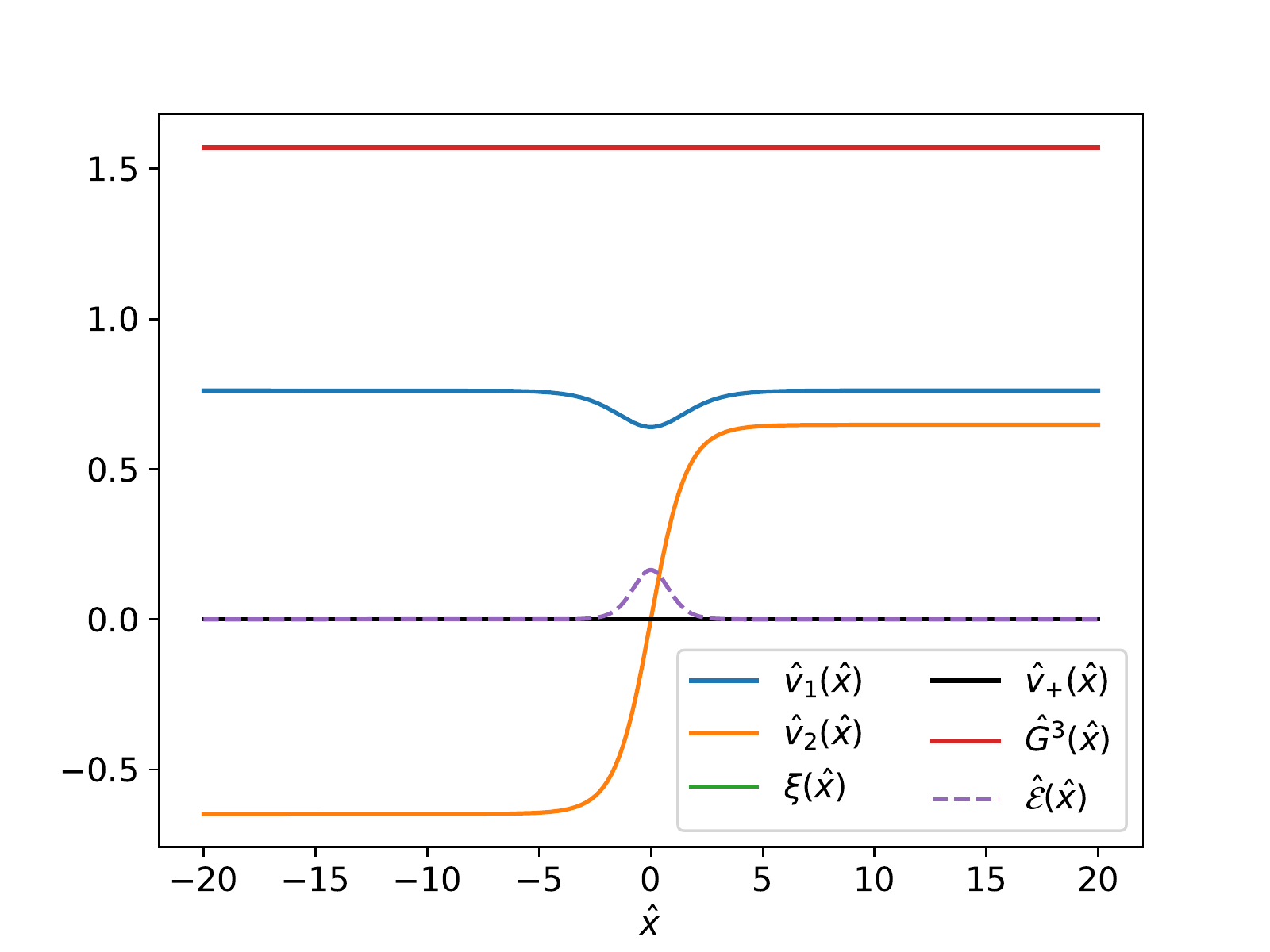}}
\subfloat[]{\includegraphics[width=0.5\textwidth]{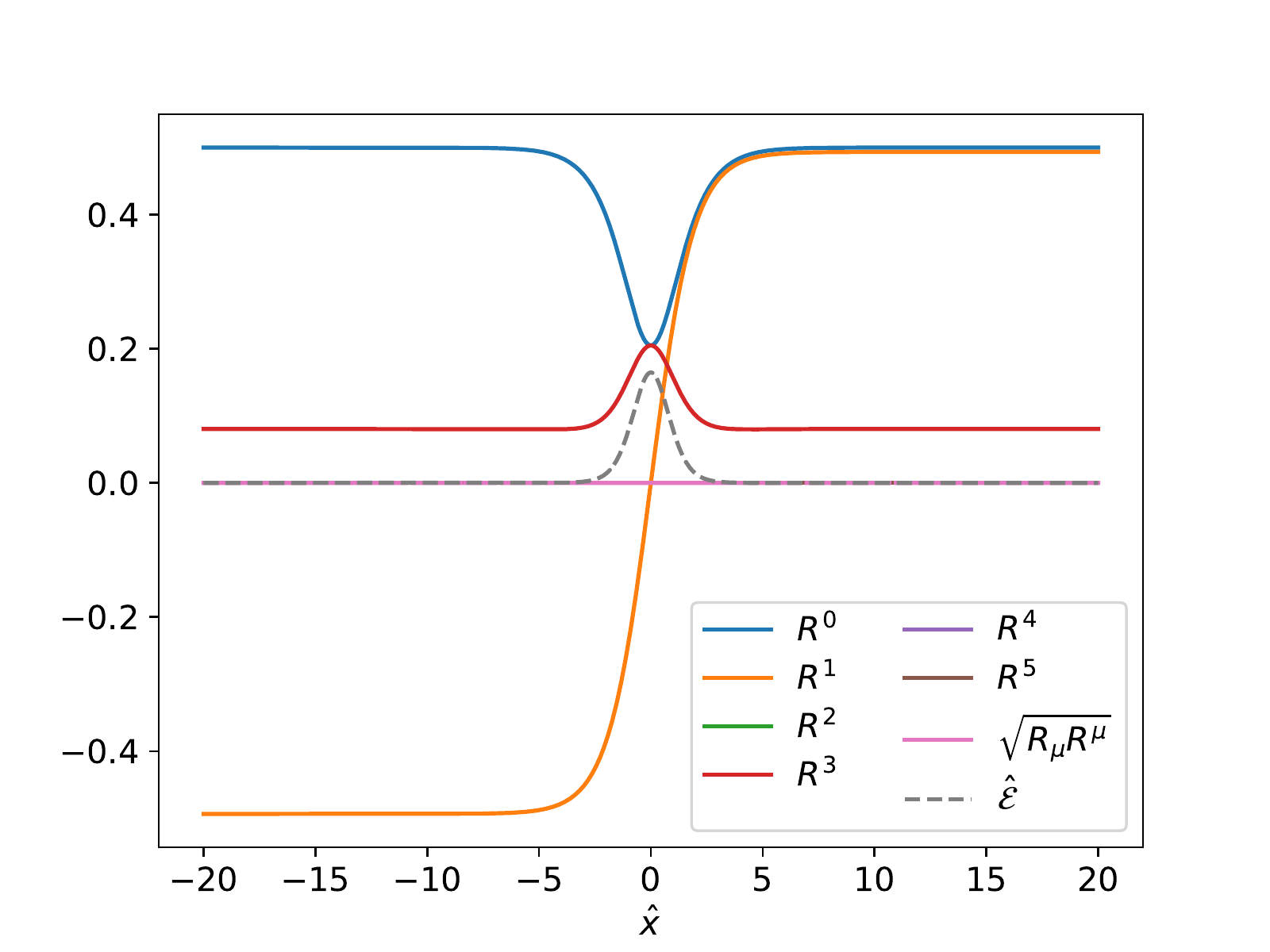}}
\hfill
\subfloat[]{\includegraphics[width=0.5\textwidth]{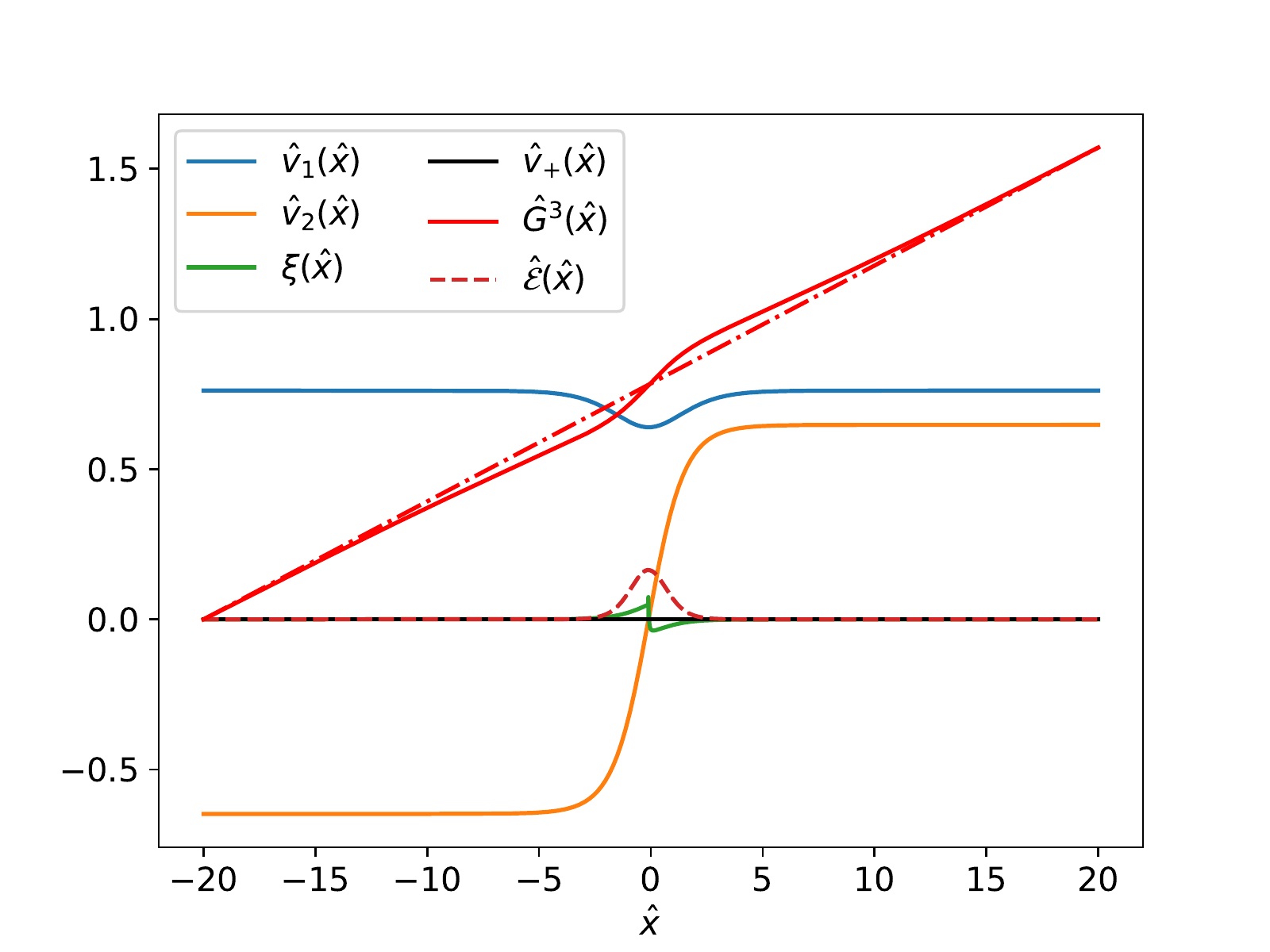}}
\subfloat[]{\includegraphics[width=0.5\textwidth]{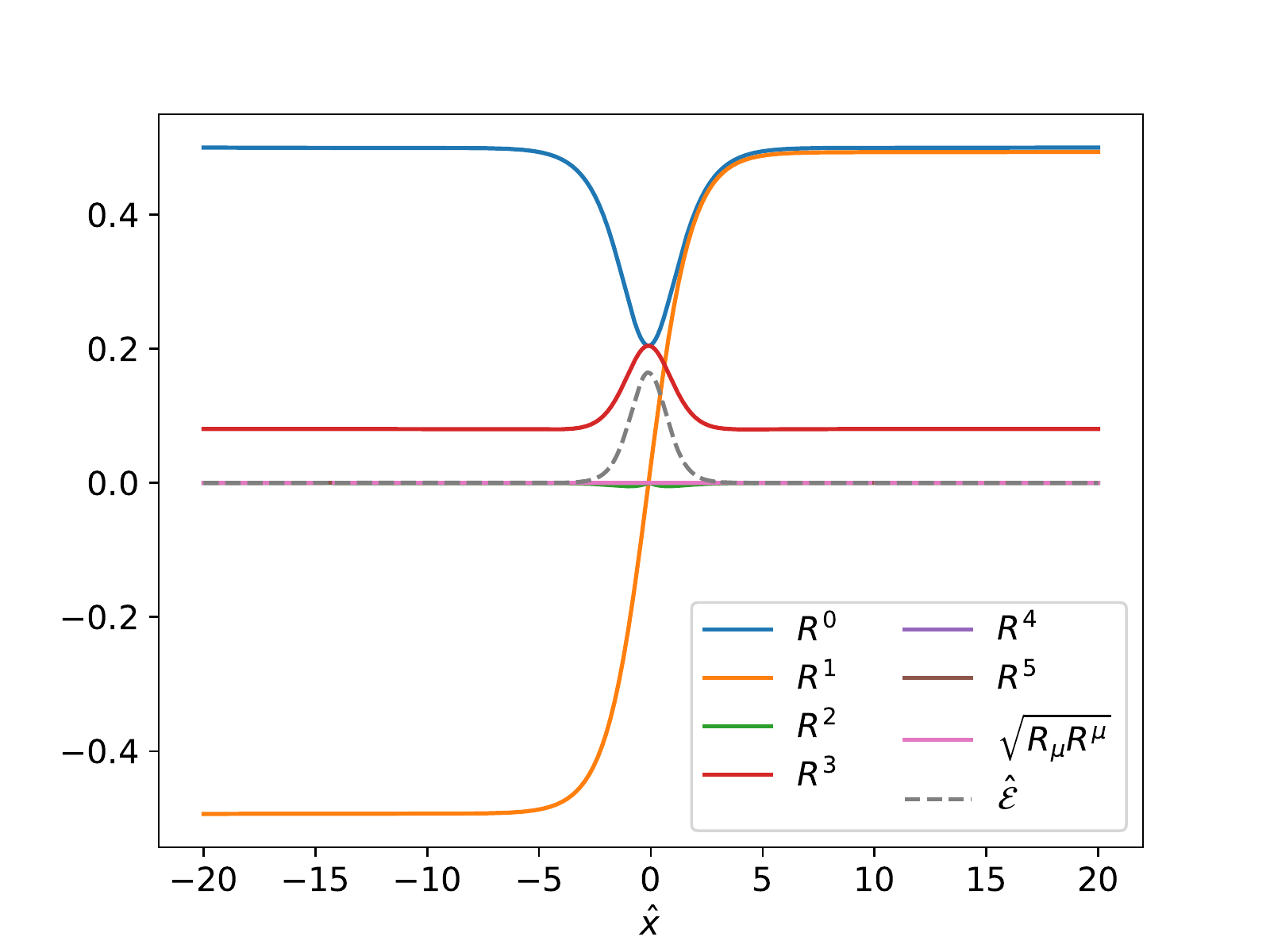}} }
\caption{Numerical evaluation of the kink parameters for different boundary conditions
  on $\widehat{G}^{3}(\hat{x})$: {\bf (a)}~$\widehat{G}^{3}(\hat{x})=\pi /2$ at both boundaries, {\bf (b)}~$R$-field profiles with $\widehat{G}^{3}(\hat{x})= \pi /2$ at both boundaries, {\bf (c)}~$\widehat{G}^{3}(\hat{x})=0$ at the LH boundary and $\widehat{G}^{3}(\hat{x})=\pi /2$ at the RH boundary. The dash-dotted line through the boundary points is shown for comparison. {\bf (d)}~Profiles in the $R$-field space with $\widehat{G}^{3}(\hat{x})=0$ at the LH boundary and $\widehat{G}^{3}(\hat{x})=\pi /2$ at the RH boundary.}
\label{Figure6.4}
\end{figure}

From the gradient flow equation for $\widehat{G}^{3}$, we obtain the constraining relation,
\begin{equation}
\label{eq3.22}
\frac{d\widehat{G}^{3}}{dx}\, =\, \frac{2v_{2}^{2}}{v_{1}^{2}+v_{2}^{2}+v_{+}^{2}}\,\frac{d\xi}{dx}\; .
\end{equation}
Equation~\eqref{eq3.22} tells us that if we have $d\widehat{G}^{3}/dx\neq 0$ for some localised and finite interval of~$x$ as $L\to\infty$, we should then have $d\xi (x)/dx\neq 0$ and $v_{2}(x) \neq 0$ on this correlated $x$-interval. Hence, we expect that the resulting kink solution be CP-violating, but electrically neutral.

In Figure \ref{Figure6.4}, we show our results obtained by the gradient flow approach.
Figure~\ref{Figure6.4}(a) shows the result using the boundary conditions $\widehat{G}^{3}= \pi/2$ at both the LH and RH boundaries, with the corresponding profiles in the $R$-space shown in Figure \ref{Figure6.4}(b). Figure \ref{Figure6.4}(c) shows the results obtained by imposing $\widehat{G}^{3}=0$ at the LH boundary and $\widehat{G}^{3}=\pi/2$ at the RH boundary, with the corresponding  $x$-profiles in the $R$-space depicted in Figure~\ref{Figure6.4}(d). Like in the $\theta$- and $\widehat{G}^{1}$-scenarios, no violation of the vacuum neutrality condition is observed, independently of whether symmetric or asymmetric boundary conditions are applied to $\widehat{G}^{3}$. If asymmetric boundary conditions are used, however, we find a non-zero $\xi(x)$ close to the kink at the origin. The latter implies that the 2HDM kink solution violates CP, as can be analytically inferred from~\eqref{eq3.22}.

\section{Discussion}
\label{Chapter7}

It is well known that the Two Higgs Doublet Model (2HDM) may possess accidental discrete\- symmetries like CP or $Z_2$ symmetry, which can be utilised to explain the origin${}$ of CP violation in nature~\cite{TDLee}, or forbid tree-level flavour-changing neutral currents in Higgs interactions~\cite{Glashow:1976nt}. However, the spontaneous breakdown of such symmetries during the electro\-weak phase transition in the early universe can give rise to the formation of domain walls that may have detrimental effects on the cosmic evolution of the early universe.

In this paper, we have extended a recent work on this topic~\cite{SDW} and studied in more detail the analytic properties of charged and CP-violating kink solutions in a $Z_{2}$-symmetric 2HDM. To do so, we have first derived the complete set of equations of motion that describe\- the 1D spatial profiles, not only of the 2HDM vacuum parameters alone as done in~\cite{SDW}, but also of the would-be Goldstone bosons $G^{1,2,3}$ associated with the SM $W^\pm$ and $Z$ bosons, and the mode $\theta$ corresponding to a would-be massive photon emerging from a possible spontaneous breakdown of the~U(1)$_{\rm em}$ group of electromagnetism. These equations are then solved numerically using the gradient flow method, and the results of our analysis are presented in the non-linear and $R$-space field representations. In particular, by virtue of~\eqref{eq43}, we have analytically demonstrated how an electrically charged profile can arise in 1D kink solutions\- when asymmetric boundary conditions are imposed on $G^{2}$ at spatial infinities, such that $G^{(2)}(-\infty) \neq G^{(2)} (+\infty)$. If similar~asymmetric boundary conditions are selected for the longitudinal mode~$\theta$ or the Goldstone mode~$G^{3}$, we have shown how the derived kink solutions obeying the constraining equations~\eqref{eq3.10} and~\eqref{eq3.22} exhibit CP violation, while preserving electric charge. These findings were corroborated by our numerical analysis\- based on the gradient flow method. We note, however, that the kink solutions obtained when asymmetric boundary values were imposed on $G^1$ do appear to respect both CP and electric charge, at least at the level of our numerical accuracy. Hence, where comparisons were possible, our results agree well with the numerical simulations carried out in~\cite{SDW}.

It is important to stress here that the total finite energy of the 2HDM kink solutions depends on the boundary conditions on the Goldstone modes $\theta$ and $G^a$ at spatial infinities. It gets higher when these conditions are asymmetric thereby triggering electric charge or CP non-conservation.  In a fashion very analogous to the well-known Lee's mechanism of spontaneous CP violation~\cite{TDLee} caused by the zero-energy vacuum state of a CP-invariant 2HDM potential, charged and CP-violating kink configurations of the ground state will now induce electric charge and CP violation for the constrained $Z_2$-symmetric 2HDM potential. It should be appreciated here the fact that spontaneous CP violation is not possible in an exact $Z_2$-symmetric 2HDM potential~\cite{Branco:2011iw}, but only through topological kink configurations as discussed in this paper.

Electrically charged and CP-violating domain walls may have a number of cosmological implications while they are decaying in the early universe. They will interact with photons and other light charged particles affecting the CMB spectrum, and in certain instances, they may even affect gravitational wave detectors, such as LIGO~\cite{Stadnik:2014tta,Jaeckel:2016jlh,Grote:2019uvn,Khoze:2021uim}.  Since the photons can become massive inside the domain walls, the latter will acquire superconducting${}$ properties as other topological defects~\cite{Witten:1984eb}. Moreover, photons will be reflected by the walls for sufficiently low frequencies below the symmetry-breaking scale~\cite{Battye:2021dyq}.  In a decaying domain-wall scenario, charge violation of the kink solution may lead to conversions of electrons or muons into neutrinos and photons or other gauge bosons~\cite{Lahanas:1998wf}, potentially modifying the relic abundances of the latter particles. Even though numerical simulations and estimates are bound to be highly model-dependent, the minimal $Z_2$-symmetric 2HDM that we have been considering here is certainly an archetypal framework for conducting realistic studies.

We note that electrically charged and CP-violating domain walls do not only occur in the $Z_2$-symmetric 2HDM under study, but they can also be a generic feature of many SM extensions for which domain walls happen to carry electroweak or other charges of gauge groups that mix with the U(1)$_{\rm em}$ group. For instance, this can be the case for some breaking patterns of Grand Unified Theories like SO(10)~\cite{Georgi:1974my, Fritzsch:1974nn}, which can go to the SM gauge group via the Pati--Salam~(PS) subgroup~\cite{Pati:1974yy}:~$\text{SO}(10) \to \text{SU}(4)_\text{PS} \times \text{SU}(2)_L \times \text{SU}(2)_R \to \text{SU}(3)_c\times \text{SU}(2)_L \times \text{SU}(2)_R \times \text{U}(1)_{B-L}\times {\cal C} \to \text{SU}(3)_c\times \text{SU}(2)_L \times \text{U}(1)_Y \to \text{SU}(3)_c\times \text{U}(1)_{\text{em}}$.  In~this breaking pattern~\cite{Kibble:1982ae,Kibble:1982dd}, ${\cal C}$ is a discrete charge conjugation symmetry, which reflects the symmetry of the theory under the interchange of left and right chiral fields belonging to the $\text{SU}(2)_L$ and $\text{SU}(2)_R$ groups, respectively, followed by a charge-conjugation of their representations. An alternative left-right symmetric theory with a similar discrete symmetry (not embeddable in SO(10)) was given in~\cite{Senjanovic:1975rk}.
Thus, an asymmetric spontaneous symmetry breaking of the SU(2)$_L$ and SU(2)$_R$ gauge groups  through hierarchical VEVs will break ${\cal C}$ spontaneously, producing a system of
domain walls bounded by strings~\cite{Kibble:1982ae,Kibble:1982dd}. Hence, the spontaneous breaking of ${\cal C}$ will potentially give rise to electrically charged and CP-violating domain walls, even in the absence of any explicit or spontaneous source of CP violation.

Finally, it would be interesting to explore whether other topological defects, such as cosmic strings and monopoles, may also carry electric charge,  or whether they can localise a non-trivial CP-violating phase close to their origin.  An~ultimate\- goal of such studies would be to understand the role that the so-generated CP-violating topological defects can play in the dynamics of electroweak baryogenesis in the 2HDM and beyond.
\vspace{-3mm}

\subsection*{Acknowledgements}
\vspace{-3mm}
We thank Richard Battye for useful discussions.  The work of AP is supported in part by the Lancaster-Manchester-Sheffield Consortium for Fundamental Physics, under STFC research grant ST/T001038/1.

\vfill\eject
\appendix
\section{Gradient Flow Technique in the $Z_{2}$-symmetric 2HDM}
\label{AppendixA}

In order to solve rather complex time-independent equations of motion that give rise to stable topological defects in extensions of the SM, like the 2HDM, we must rely on numerical methods. One such convenient method is the so-called gradient flow technique, which enables one to numerically solve a set of coupled second-order differential equations, with well defined Neumann or Dirichlet initial conditions~\cite{Battye:2011jj,Brawn:2012mgn}. We have applied the gradient flow technique to obtain numerical solutions for one-dimensional (1D) topological kink configurations.

In detail, the 1D energy density $\mathcal{E}$ of the $Z_{2}$-symmetric 2HDM is given
by the $\{00\}$ component of the energy stress tensor, $T^{00}$,  i.e.
\begin{equation}\label{eq5.14}
\mathcal{E}(\Phi_{1}, \Phi_{2})=\frac{\D\Phi_{1}^{\dagger}}{\D x}\frac{d\Phi_{1}}{dx}+\frac{\D\Phi_{2}^{\dagger}}{\D x}\frac{d\Phi_{2}}{dx}+V(\Phi_{1}, \Phi_{2})+V_{0}\:,
\end{equation}
where $V(\Phi_{1}, \Phi_{2})$ is the $Z_{2}$-symmetric potential given in~\eqref{eq16} and $V_{0}$ is a constant introduced here in order to shift the minimum energy density to zero, such that $\mathcal{E}$ is non-negative for all~$x$. We use the gradient flow technique to find solutions that minimize the total energy of the system, $E=\int\D x\:\mathcal{E}(\Phi_{1}, \Phi_{2})$. In three spatial dimensions, this represents the energy per unit area of the system. We introduce a fictitious time parameter $t$, so that the ground-stated functions within the Higgs doublets, collectively denoted here as $f_i$, become $t$-dependent, 
\begin{equation}
f_{i}\, =\, f_{i}(x, t)\; .
\end{equation}
Since we wish our field solutions to minimize the total energy $E$ of the system,  we set
\begin{equation}\label{eq5}
\dot{f_{i}}=-\frac{\delta E}{\delta f_{i}}\; .
\end{equation}
We introduced in~\eqref{eq5} a negative sign before the functional derivative, because we want the fields to evolve in a way such that the energy decreases and eventually reaches a minimum.

In our numerical analysis, we have appropriately redefined the length~$x$, the vacuum para\-meters~$v_{1,2,+}$ and the energy density $\mathcal{E}$, so as to become dimensionless,
\begin{equation}
\hat{x}\, \equiv\, M_{h}x\:,\qquad \hat{v}_{1,2,+}\, \equiv\, \frac{v_{1,2,+}}{v_{\textrm{SM}}}\:,\qquad \hat{\mathcal{E}}\, \equiv\, \frac{\mathcal{E}}{M_{h}^{2}v_{\textrm{SM}}^{2}}\:,
\end{equation}
where $M_{h}=125\:\textrm{GeV}$ is the value used for the SM Higgs mass and $v_{\textrm{SM}}= 246\:\textrm{GeV}$ is the VEV assumed for the SM Higgs field. These values coincide with their central values as determined by experiment~\cite{Aad}. Finally, we have redefined all relevant kinematic parameters of the $Z_2$-symmetric 2HDM to facilitate the rescaling of the energy density: \begin{equation}\label{eq5.3.14}
  \hat{\mu}_{1}^{2}=\frac{\mu_{1}^{2}}{M_{h}^{2}}\:,\qquad \hat{\mu}_{2}^{2}=\frac{\mu_{2}^{2}}{M_{h}^{2}}\:,\qquad
  \hat{\lambda}_{i}=\frac{\lambda_{i}v_{\textrm{SM}}^{2}}{M_{h}^{2}}\,,
\end{equation}
with $i=1, 2, 3, 4, 5$.

\end{document}